\documentclass[aps,groupedaddress,twocolumn,superscriptaddress]{revtex4-2}

\usepackage{graphicx}%
\usepackage{multirow}%
\usepackage{amsmath,amssymb,amsfonts}%
\usepackage[citecolor=blue,linkcolor=blue,urlcolor=blue]{hyperref}
\usepackage[ocgcolorlinks]{ocgx2}
\usepackage{sidecap}
\usepackage{bm}
\usepackage{makecell}

\newcommand{\penta}{\textsc{Pentatrap}\ }
\newcommand{\pentano}{\textsc{Pentatrap}}

\newcommand{\rtwo}{\ensuremath{\langle r^2 \rangle}}
\newcommand{\rfour}{\ensuremath{\langle r^4 \rangle}}
\newcommand{\drtwo}{\ensuremath{\delta \langle r^2 \rangle}}
\newcommand{\drfour}{\ensuremath{\delta \langle r^4 \rangle}}

\newcommand{\ket}[1]{\ensuremath{| #1 \rangle}}
\newcommand{\elem}[2]{\ensuremath{{}^{#2}\text{#1}}}

\newcommand{\orbital}[3]{\ensuremath{#1 #2_{#3 /2}}}
\newcommand{\dnnlogo}{\ensuremath{\Delta\text{N}^2\text{LO}_{\text{GO}}}}
\newcommand{\fm}{\ensuremath{\text{fm}}}
\newcommand{\Lm}{\mathbf{\Lambda_-}}
\newcommand{\Lp}{\mathbf{\Lambda_+}}

\newcommand{\mIS}[1]{\ensuremath{\tilde{#1}}}
\newcommand{\nuclearmodelone}{\ensuremath{1.8/2.0\:(\text{EM}),\text{VS1}}}
\newcommand{\nuclearmodeltwo}{\ensuremath{1.8/2.0\:(\text{EM}),\text{VS2}}}
\newcommand{\nuclearmodelthree}{\ensuremath{\dnnlogo{},\text{VS1}}}

\newcommand{\nocontentsline}[3]{}
\newcommand{\tocless}[2]{\bgroup\let\addcontentsline=\nocontentsline#1{#2}\egroup}

\begin{document}

\preprint{APS/123-QED}

\title{Probing new bosons and nuclear structure with ytterbium isotope shifts}

\author{Menno Door}
\thanks{These authors contributed equally to this work. Corresponding authors: menno.door@mpi-hd.mpg.de, chih-han.yeh@ptb.de}
\affiliation{Max-Planck-Institut für Kernphysik, Saupfercheckweg 1, 69117 Heidelberg, Germany}
\affiliation{Heidelberg University, Grabengasse 1, 69117 Heidelberg, Germany}
\author{Chih-Han Yeh}
\thanks{These authors contributed equally to this work. Corresponding authors: menno.door@mpi-hd.mpg.de, chih-han.yeh@ptb.de}
\affiliation{Physikalisch-Technische Bundesanstalt, Bundesallee 100, 38116 Braunschweig, Germany}
\author{Matthias Heinz}
\thanks{Present address: National Center for Computational Sciences, Oak Ridge National Laboratory, Oak Ridge, TN 37831, USA}
\affiliation{Department of Physics, Technische Universität Darmstadt, Darmstadt, 64289, Germany}
\affiliation{ExtreMe Matter Institute EMMI, GSI Helmholtzzentrum für Schwerionenforschung GmbH, Darmstadt, 64291, Germany}
\affiliation{Max-Planck-Institut für Kernphysik, Saupfercheckweg 1, 69117 Heidelberg, Germany}
\author{Fiona Kirk}
\affiliation{Physikalisch-Technische Bundesanstalt, Bundesallee 100, 38116 Braunschweig, Germany}
\affiliation{Institut für Theoretische Physik, Leibniz Universität Hannover, Appelstraße 2, 30167 Hannover, Germany}
\author{Chunhai Lyu}
\affiliation{Max-Planck-Institut für Kernphysik, Saupfercheckweg 1, 69117 Heidelberg, Germany}
\author{Takayuki Miyagi}
\affiliation{Department of Physics, Technische Universität Darmstadt, Darmstadt, 64289, Germany}
\affiliation{ExtreMe Matter Institute EMMI, GSI Helmholtzzentrum für Schwerionenforschung GmbH, Darmstadt, 64291, Germany}
\affiliation{Max-Planck-Institut für Kernphysik, Saupfercheckweg 1, 69117 Heidelberg, Germany}
\author{Julian C. Berengut}
\affiliation{School of Physics, University of New South Wales, Sydney, New South Wales 2052, Australia}
\author{Jacek Bieroń}
\affiliation{Institute of Theoretical Physics, Jagiellonian University, Kraków, 30-348, Poland}
\author{Klaus Blaum}
\affiliation{Max-Planck-Institut für Kernphysik, Saupfercheckweg 1, 69117 Heidelberg, Germany}
\author{Laura S. Dreissen}
\affiliation{Physikalisch-Technische Bundesanstalt, Bundesallee 100, 38116 Braunschweig, Germany}
\affiliation{Department of Physics and Astronomy, LaserLab, Vrije Universiteit Amsterdam, De Boelelaan 1081, Amsterdam, 1081 HV, The Netherlands}
\author{Sergey Eliseev}
\affiliation{Max-Planck-Institut für Kernphysik, Saupfercheckweg 1, 69117 Heidelberg, Germany}
\author{Pavel Filianin}
\affiliation{Max-Planck-Institut für Kernphysik, Saupfercheckweg 1, 69117 Heidelberg, Germany}
\author{Melina Filzinger}
\affiliation{Physikalisch-Technische Bundesanstalt, Bundesallee 100, 38116 Braunschweig, Germany}
\author{Elina Fuchs}
\affiliation{Physikalisch-Technische Bundesanstalt, Bundesallee 100, 38116 Braunschweig, Germany}
\affiliation{Institut für Theoretische Physik, Leibniz Universität Hannover, Appelstraße 2, 30167 Hannover, Germany}
\author{Henning A. Fürst}
\affiliation{Physikalisch-Technische Bundesanstalt, Bundesallee 100, 38116 Braunschweig, Germany}
\affiliation{Institut für Quantenoptik, Leibniz Universität Hannover, Welfengarten 1, 30167 Hannover, Germany}
\author{Gediminas Gaigalas}
\affiliation{Institute of Theoretical Physics and Astronomy, Vilnius University, Vilnius, 10222, Lithuania}
\author{Zoltán Harman}
\affiliation{Max-Planck-Institut für Kernphysik, Saupfercheckweg 1, 69117 Heidelberg, Germany}
\author{Jost Herkenhoff}
\affiliation{Max-Planck-Institut für Kernphysik, Saupfercheckweg 1, 69117 Heidelberg, Germany}
\author{Nils Huntemann}
\affiliation{Physikalisch-Technische Bundesanstalt, Bundesallee 100, 38116 Braunschweig, Germany}
\author{Christoph H. Keitel}
\affiliation{Max-Planck-Institut für Kernphysik, Saupfercheckweg 1, 69117 Heidelberg, Germany}
\author{Kathrin Kromer}
\affiliation{Max-Planck-Institut für Kernphysik, Saupfercheckweg 1, 69117 Heidelberg, Germany}
\author{Daniel Lange}
\affiliation{Max-Planck-Institut für Kernphysik, Saupfercheckweg 1, 69117 Heidelberg, Germany}
\affiliation{Heidelberg University, Grabengasse 1, 69117 Heidelberg, Germany}
\author{Alexander Rischka}
\affiliation{Max-Planck-Institut für Kernphysik, Saupfercheckweg 1, 69117 Heidelberg, Germany}
\author{Christoph Schweiger}
\affiliation{Max-Planck-Institut für Kernphysik, Saupfercheckweg 1, 69117 Heidelberg, Germany}
\author{Achim Schwenk}
\affiliation{Department of Physics, Technische Universität Darmstadt, Darmstadt, 64289, Germany}
\affiliation{ExtreMe Matter Institute EMMI, GSI Helmholtzzentrum für Schwerionenforschung GmbH, Darmstadt, 64291, Germany}
\affiliation{Max-Planck-Institut für Kernphysik, Saupfercheckweg 1, 69117 Heidelberg, Germany}
\author{Noritaka Shimizu}
\affiliation{Center for Computational Sciences, University of Tsukuba, Ibaraki, 305-8577, Japan}
\author{Tanja E. Mehlstäubler}
\affiliation{Physikalisch-Technische Bundesanstalt, Bundesallee 100, 38116 Braunschweig, Germany}
\affiliation{Institut für Quantenoptik, Leibniz Universität Hannover, Welfengarten 1, 30167 Hannover, Germany}
\affiliation{Laboratorium für Nano- und Quantenengineering, Leibniz Universität Hannover, Schneiderberg 39, 30167 Hannover, Germany}

\begin{abstract}
In this Letter, we present mass-ratio measurements on highly charged Yb$^{42+}$ ions with a precision of $4\times 10^{-12}$ and isotope-shift measurements on Yb$^{+}$ on the $^{2}$S$_{1/2}$ → $^{2}$D$_{5/2}$ and $^{2}$S$_{1/2}$ → $^{2}$F$_{7/2}$ transitions with a precision of $4\times 10^{-9}$ for the isotopes $^{168,170,172,174,176}$Yb. We present a new method that allows us to extract higher-order changes in the nuclear charge distribution along the Yb isotope chain, benchmarking \textit{ab initio} nuclear structure calculations.
Additionally, we perform a King plot analysis to set bounds on a fifth force in the keV$/c^2$ to MeV$/c^2$ range coupling to electrons and neutrons.
\end{abstract}

\keywords{Isotope-shift spectroscopy, fifth forces, new bosons, high-precision mass spectrometry, \textit{ab initio} atomic and nuclear structure calculations, nuclear charge distributions}

\maketitle

Theories beyond the Standard Model (SM) of particle physics are typically probed by high-energy colliders or astrophysical and cosmological observations. Competitive complementary tests can be performed with high-precision atomic and molecular physics experiments at low energies~\cite{safronova2018search}. In particular, isotope-shift spectroscopy, commonly used to study nuclear charge radii in exotic isotopes~\cite{GarciaRuiz:2016ohj}, is also sensitive to shifts in atomic energy levels induced by hypothetical new bosons that mediate an additional interaction between neutrons and electrons~\cite{Delaunay2017probing,berengut2018probing}. Such measurements can be analyzed via the King-plot method, where different atomic transitions are combined in such a way that common nuclear and atomic uncertainties are eliminated. Deviations from the linearity of the King plot indicate effects from new physics or higher-order atomic and nuclear structure. This powerful technique has been successfully used to put bounds on physics beyond the SM for example with isotope shifts measured in ytterbium~\cite{Counts2020evidence, Allehabi:2020xgf, Hur2022evidence}. With increasing precision of the frequency measurements, the uncertainties of the nuclear masses~\cite{Counts2020evidence, nesterenko2020_yb168mass} become a limiting factor for distinguishing between higher-order SM effects and new physics. 

In this Letter, we present high-precision mass-ratio and isotope-shift measurements of five stable, spinless ytterbium isotopes. Both the mass spectrometry and the isotope-shift spectroscopy are up to two orders of magnitude more precise than previous measurements~\cite{Hur2022evidence, rana2012_ybmasses,nesterenko2020_yb168mass}. The isotope mass ratios are determined using highly charged Yb ions in the Penning-trap mass spectrometer \pentano~\cite{repp2012_pentatrapscheme}, reaching a relative precision of a few $10^{-12}$ contributing to a relative uncertainty of $10^{-10}$ for the mass-normalized isotope shifts. The isotope-shift spectroscopy is performed on $\mathrm{Yb}^{+}$ on the $^{2}\!S_{1\!/\!2}\!\rightarrow\!{}^{2}\!D_{5\!/\!2}$ and the $^{2}\!S_{1\!/\!2}\!\rightarrow\!{}^{2}\!F_{7\!/\!2}$ transitions with a relative precision as low as $10^{-9}$. 
Our results deviate significantly from some former mass-ratio and isotope-shift measurements.
Using our improved data, we construct a generalized King plot~\cite{Berengut:2020itu} and extract a competitive spectroscopic exclusion bound on the coupling strength of potential new bosons to electrons and neutrons.

Combining these precise measurements with atomic structure calculations allows us to investigate higher-order nuclear structure effects in Yb isotopes~\cite{Allehabi:2020xgf} and extract changes in the quartic charge radius $\drfour$ along the isotopic chain, providing a new window into nuclear deformation.
Building on advances in \textit{ab initio} nuclear structure calculations~\cite{Hergert2016imsrg,Stroberg2019vsimsrg,Hergert:2020bxy,Miyagi2022no2b3n}, we provide a first microscopic description of Yb nuclei starting from chiral effective field theory interactions~\cite{Epelbaum:2008ga} based on quantum chromodynamics. This method can provide direct insights into the evolution of nuclear charge distributions along isotopic chains towards exotic, neutron-rich nuclei.

{\it Theoretical framework.--}
An isotope shift is the difference in the frequencies of a given atomic transition in two different isotopes of the same element.
Here, we consider the $^{2}\!S_{1\!/\!2}\!\rightarrow\!{}^{2}\!D_{5\!/\!2}$ electric quadrupole and the highly forbidden $^{2}\!S_{1\!/\!2}\!\rightarrow\!{}^{2}\!F_{7\!/\!2}$ electric octupole transitions, denoted as $\alpha$ and $\gamma$, in singly charged Yb$^+$ ions. We consider five stable, even Yb isotopes with mass numbers $A\in\{168, 170, 172, 174, 176\}$ containing four neighboring isotope pairs $(A,A')$ with $A'=A+2$. The corresponding isotope shifts $\nu_\alpha ^{A,A'} = \nu_\alpha^A-\nu_\alpha^{A'}$ make up the entries of a four-component vector $\boldsymbol{\nu}_\alpha$ for transition $\alpha$ (and similarly for $\gamma$). 
This vector $\boldsymbol{\nu}_\alpha$ can be written as a linear combination of the field and mass shifts~\cite{king1963comments}, additional higher-order SM-based shifts, and a term induced by the interaction with the proposed boson. Each of these terms can be decomposed into an electronic factor (with subscript $\alpha$) and a vector encoding the nuclear structure:
\begin{equation}
    \begin{aligned}
        \boldsymbol{\nu}_\alpha=& 
        F_\alpha \boldsymbol{\delta}\langle r^{2}\rangle 
        + K_\alpha \boldsymbol{w} 
        + G_{\alpha}^{(2)} \boldsymbol{\delta}\langle r^{2}\rangle^{2} 
        + G_{\alpha}^{(4)} \boldsymbol{\delta}\langle r^{4}\rangle \\
        &+ \frac{\alpha_{\mathrm{NP}}}{\alpha_{\mathrm{EM}}}D_{\alpha}\boldsymbol{h} + \ldots\,.
    \end{aligned}
    \label{eq:isotopeshift_all}
\end{equation}
$F$, $K$, $G^{(2)}$, $G^{(4)}$, and $D$ are transition-dependent factors for the multiplicative electronic contribution to the field shift, mass shift, quadratic field shift, quartic shift, and a shift induced by a new boson, respectively. 
The components of $\boldsymbol{w}$ are $w^{A,A'} = m_{172}/m_{A} - m_{172}/m_{A'} = 1/\eta_A-1/\eta_{A'}$, the inverse nuclear mass differences of the isotopes $A$ and $A'$ with respect to the nuclear mass of \elem{Yb}{172}. $\boldsymbol{\delta}\langle r^{n}\rangle$ is the four-vector with elements $\delta\langle r^{n}\rangle^{A,A'} = \langle r^{n}\rangle^{A} - \langle r^{n}\rangle^{A'}$, the differences between the $n$-th moments of the nuclear charge distributions.
$\boldsymbol{\delta}\langle r^{2}\rangle^2$ has the components $(\drtwo{}^2)^{A,A'} = (\drtwo{}^{A,176})^2 - (\drtwo{}^{A',176})^2$
constructed from squared radius differences.
Additional higher-order SM contributions may also contribute to Eq.~\eqref{eq:isotopeshift_all} at a given experimental accuracy.
$\alpha_{\mathrm{NP}}=(-1)^{s+1}y_ny_e/(4\pi\hbar c)$ is the product of the coupling constants $y_n$ and $y_e$ of the new boson (with mass $m_\phi$ and spin $s$) to the neutron and electron. It enters the Yukawa potential $V_{\text{ne}}(r)=\hbar c\cdot\alpha_{\mathrm{NP}}\cdot\exp(-r m_\phi c/\hbar)/r$~\cite{berengut2018probing} generated by the new boson. We normalize $\alpha_{\mathrm{NP}}$ by the fine-structure constant $\alpha_{\mathrm{EM}}=e^2/(4\pi \hbar c)$, where $e$ is the elementary charge. $\boldsymbol{h}=-(2,2,2,2)$ is the vector of neutron number differences for the neighboring isotope pairs.

Usually, only the first two terms in Eq.~\eqref{eq:isotopeshift_all} contribute significantly. With isotope-shift measurements for two transitions $\alpha$ and $\gamma$, one can eliminate $\boldsymbol{\delta}\langle r^{2}\rangle$ from Eq.~\eqref{eq:isotopeshift_all} so that the first two terms yield the linear King relation~\cite{king1963comments,berengut2018probing}
\begin{align}
    \mIS{\boldsymbol{\nu}}_{\gamma}\,\approx\,& 
    F_{\gamma\alpha}\mIS{\boldsymbol{\nu}}_{\alpha} + K_{\gamma\alpha}\boldsymbol{1}\,,
    \label{eq:kingrelation_vector}
\end{align}
with the mass-normalized $\mIS{\boldsymbol{\nu}}_{\gamma} = \boldsymbol{\nu}_{\gamma}/\boldsymbol{w}$, 
$\boldsymbol{1}= (1,1,1,1)$, $F_{\gamma\alpha} = F_\gamma/F_\alpha$, and $K_{\gamma\alpha} = K_\gamma-F_{\gamma\alpha}K_\alpha$. 
Deviations from Eq.~\eqref{eq:kingrelation_vector} indicate the presence of additional terms beyond the leading-order field and mass shifts, as in Eq.~\eqref{eq:isotopeshift_all}.

{\it Experimental results.--}

\begin{figure*}[ht!]
    \centering
    \includegraphics[width=\linewidth]{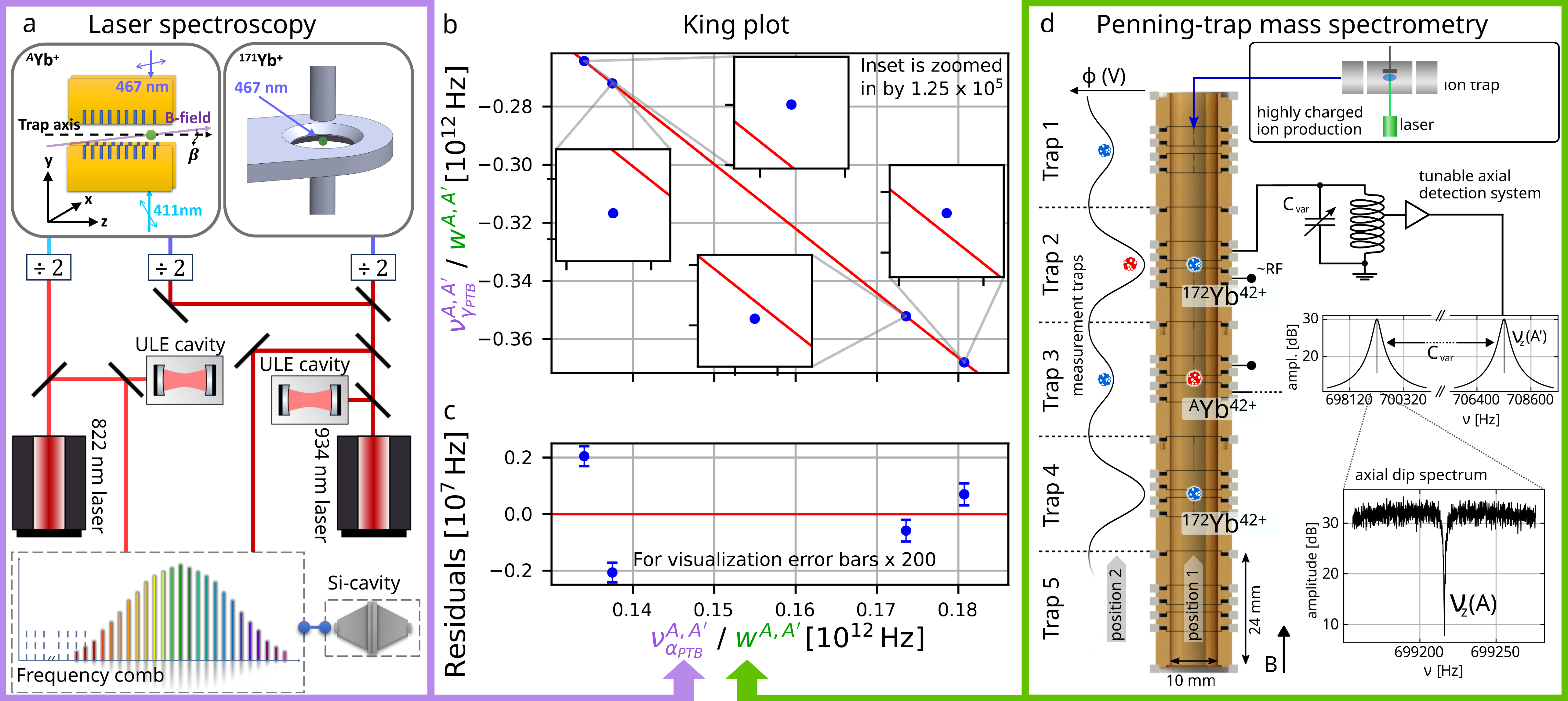}
    \caption{Scheme of experimental setups and the King-plot analysis. \textbf{a} Laser spectroscopy setup for the optical frequency ratio measurements of the transitions near 411\,nm ($\alpha$) and 467\,nm ($\gamma$). The laser fields with wavelengths near 822\,nm and 934\,nm are first stabilized to an ultra-low expansion (ULE) cavity for short-term frequency stability and then locked to a cryogenic silicon cavity (Si-cavity). The second harmonic conversions of the lasers are focused to interrogate the ions. 
    \textbf{b} King plot of the isotope shifts of the $\gamma$ transition with respect to the isotope shifts of the $\alpha$ transition normalized with the inverse mass-ratio difference $w^{A,A'}$, with insets magnified by a factor of $1.25\times10^{5}$. \textbf{c} Residuals of the linear fit in the King plot. For visibility, the uncertainties on the residuals are multiplied by a factor of 200. \textbf{d} Penning-trap setup for the determination of cyclotron frequency ratios. Three highly charged ions (blue and red), produced by an electron beam ion trap, are transported and stored in an $A'$-$A$-$A'$ sequence. By shuttling the set of ions up and down between neighbouring traps (alternating between positions 1 and 2), cyclotron frequency ratios are determined from sequential measurements in the two measurement traps, referred to trap 2 and 3 in the figure. The tunable image-current detection system allows for the determination of the axial eigenfrequencies of different isotopes at equal charge state and equal trapping potential ~\cite{heisse2017_protonmass_varactor}. The radial eigenfrequencies are determined indirectly using the same detection system.}
    \label{fig:experimental_setup}
\end{figure*}

In Fig.~\ref{fig:experimental_setup} we present an overview over the two experimental setups as well as a King plot of the isotope shifts of the $\gamma$ transition with respect to the isotope shifts of the $\alpha$ transition normalized with the inverse mass-ratio difference $w^{A,A'}$. 

The nuclear mass ratios $\eta_A$ of ytterbium isotopes are determined from the cyclotron frequency ratios of highly charged $\text{Yb}^{42+}$ ions measured at the cryogenic Penning-trap mass spectrometer \penta~\cite{repp2012_pentatrapscheme, roux2012_pentatower, rischka2020_xenon} and their calculated electron binding energies.
The determination of the cyclotron frequencies $\nu_c$ of two isotope ions allows one to extract their ionic mass ratio via 
$\mathrm{R}_A^{\text{CF}}\!=\!\nu_{\text{c}} \left(^{172}\text{Yb}^{42+} \right)\!/ \nu_{\text{c}}\left(^{A}\text{Yb}^{42+} \right)\!=\! m \left(^{A}\text{Yb}^{42+} \right)/m \left(^{172}\text{Yb}^{42+} \right)$. Determining the free cyclotron frequency $\nu$ requires measuring all three eigenfrequencies of the trapped ion via $\nu_c^2 = \nu_+^2 + \nu_z^2 + \nu_-^2$~\cite{brown_gabriels1982_invariance}, namely the trap-modified cyclotron frequency $\nu_+$, the axial frequency $\nu_z$, and the magnetron frequency $\nu_-$.

From the calculated electron binding energies $E_{172}^{(28)}=350\,773(5)$\,eV of the 28 electrons in the $^{172}\text{Yb}^{42+}$ ion and $E_{172}^{(70)}=382\,301(16)$\,eV of the 70 electrons in the $^{172}\text{Yb}$ atom~\cite{GRASP2018}, the neutral mass $m(^{172}\text{Yb})$~\cite{rana2012_ybmasses, huang2021_ame}, and the electron mass $m_e$~\cite{tiesinga2021_codata_emass}, one can derive the necessary nuclear mass ratios $\eta_{A}$ to similar accuracies. The mass ratios are used instead of single mass values in atomic mass units, since the former are much less sensitive to the uncertainties of the binding energies and the reference mass $m(^{172}\text{Yb})$.
Their final values are given in Tab.~\ref{tab:joint_table} with relative uncertainties of $4\times 10^{-12}$, corresponding to uncertainties of $0.3$\,Hz on the isotope shifts. For comparison, previous mass determinations affected the King-plot analysis at a level of $3$--$30$\,Hz.

\begin{table*}[ht]
	\centering
	\caption{Measured values of the mass ratios and isotope shifts. Columns 2 and 3 show the Yb$^{42+}$ cyclotron frequency ratios and nuclear mass ratios of the stable, even ytterbium isotopes relative to the nuclear mass of isotope $A=172$, with the differences to Refs.~\cite{rana2012_ybmasses,nesterenko2020_yb168mass} given in Column 4. Columns 5 and 7 show the isotope shifts $\nu^{A,A+2} = \nu^{A}-\nu^{A+2}$ of the $\alpha$ transition and the $\gamma$ transition in units of Hz, with the differences to Ref.~\cite{Counts2020evidence} for $\alpha$ given in Column 6. Our isotope shifts for the $\gamma$ transition are compatible with those of Ref.~\cite{Hur2022evidence}.}
    \begin{ruledtabular}
	\begin{tabular}{lcccccc}
		$A$ & $\text{R}^{\text{CF}}_A = \nu_{c,172}/\nu_{c,A}$ &  $\eta_A=m_A/m_{172}$ & $\Delta \eta_A $ [$10^{-12}$] & $\nu_{\alpha_\mathrm{PTB}}^{A,A+2}$ [Hz] & $\Delta \nu_{\alpha}^{A,A+2}$ [Hz] &  $\nu_{\gamma_\mathrm{PTB}}^{A,A+2}$ [Hz]  \\ 
        & \\[-2.3ex] 
        \hline 
		168 & $0.976\,717\,951\,145\,(4)$ & 0.976\,715\,921\,749\,(4) & $-1890\,(780)$~\cite{nesterenko2020_yb168mass} & 2179098868.0\,(5.3) & $-62\,(210)$~\cite{Counts2020evidence} & $-4438159671.1\,(15.7)$\\ 
		170 & $0.988\,356\,814\,144\,(4)$ & 0.988\,355\,799\,258\,(4) & $-88\,(108)$~\cite{rana2012_ybmasses} & 2044851281.0\,(4.9) & $-3499\,(340)$~\cite{Counts2020evidence} & $-4149190501.1\,(15.7)$ \\ 
        172 & -- & -- & -- & 1583064149.3\,(4.8) & $-4271\,(360)$~\cite{Counts2020evidence} &  $-3132320458.1\,(15.7)$\\
		174 & $1.011\,648\,196\,817\,(4)$ &  1.011\,649\,212\,140\,(4) & $153\,(122)$~\cite{rana2012_ybmasses} & 1509053195.8\,(4.7) & $-2094\,(280)$~\cite{Counts2020evidence} &  $-2976392045.3\,(15.7)$\\ 
		176 & $1.023\,303\,526\,697\,(4)$ & 1.023\,305\,557\,965\,(4) & $68\,(173)$~\cite{rana2012_ybmasses} & -- & -- & -- \\ 
	\end{tabular}
    \end{ruledtabular}
    \label{tab:joint_table}
\end{table*}

To make use of the new mass uncertainties, we improve the uncertainties of the isotope shifts by performing absolute frequency measurements of the $\alpha$ and $\gamma$ transitions for the five isotopes. 
Singly charged ytterbium ions are trapped in a segmented, linear radio-frequency Paul trap~\cite{pyka_high-precision_2014,keller_probing_2019}. 
The excitation lasers near wavelengths of 411\,nm and 467\,nm are locked to ultra-low-expansion cavities and to a cryogenic silicon cavity~\cite{Matei2017} via a frequency comb. 

The absolute transition frequencies are obtained with optical frequency ratio measurements by referencing to the $\gamma$ transition between the $F=0$ and $F=3$ hyperfine states of the $^{171}$Yb$^{+}$ isotope~\cite{Lange2021improved}. 
From the measurements, we obtain isotope shifts shown in Tab.~\ref{tab:joint_table}, with uncertainties below 6\,Hz and 16\,Hz for the $\alpha$ and the $\gamma$ transitions, respectively.

For details on the mass-ratio measurements, spectroscopic measurements, systematic shifts, and uncertainties, see the Supplemental Material.

{\it King-plot analysis.--}
A King plot using our mass-normalized isotope-shift measurements of the 
$\gamma$ transition (denoted $\mIS{\boldsymbol{\nu}}_{\gamma_\mathrm{PTB}}$) and the $\alpha$ transition (denoted $\mIS{\boldsymbol{\nu}}_{\alpha_\mathrm{PTB}}$) is given in the End Matter.
From the King-plot analysis, we observe deviations from linearity averaging to 20.17(2)\,kHz. 

\begin{figure}[ht!]
    \centering
    \includegraphics[width=3.4in]{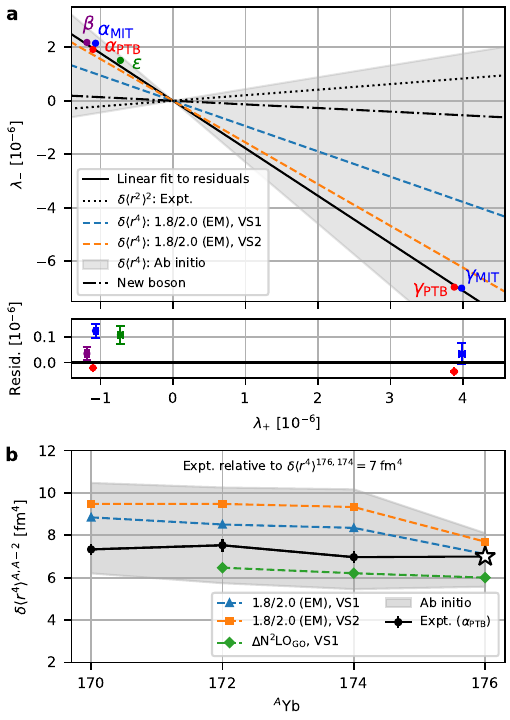}
    \caption{Nonlinearity decomposition and the extracted trend of nuclear deformation, \drfour. \textbf{a} Observed and predicted nonlinearities in the $(\lambda_+,\lambda_-)$ plane. The nonlinearities of the transitions $\alpha_\mathrm{PTB}$ (this work, see Table~\ref{tab:joint_table}), $\beta$~\cite{Counts2020evidence}, $\gamma_\mathrm{PTB}$ (this work, see Table~\ref{tab:joint_table}), $\gamma_\mathrm{MIT}$~\cite{Counts2020evidence,Hur2022evidence}, and $\epsilon$~\cite{Figueroa2022precision} with respect to the $\delta$~\cite{Ono2022observation} transition are included in the single-source linear fit (solid line) whose slope is compared to the predicted slopes for the $\drtwo^2$ (dotted), $\drfour$ (dashed lines and gray band based on predictions in \textbf{b}), and new physics (dash-dotted) nonlinearities. 
    The residuals of the single-source linear fit are shown in the lower panel.
    \textbf{b}
    Solid line: $\drfour^{A,A-2}$ values relative to $\drfour^{176,174} = 7\:\fm^4$ (star) extracted from isotope shifts of the $\alpha_{\mathrm{PTB}}$ transition using atomic theory. Dashed lines: \textit{ab initio} nuclear theory predictions [1.8/2.0 (EM), VS1 and VS2; $\Delta\text{N}^2\text{LO}_\text{GO}$, VS1]. The estimated nuclear theory uncertainties (68\% confidence interval) are given by the gray bands. 
    }
    \label{fig:NDP}
\end{figure}

To determine the origin of the nonlinearity in the King plot, we perform a nonlinearity decomposition analysis of all available (sub-)kHz-precision isotope-shift data. Apart from $\alpha_{\mathrm{PTB}}$ and $\gamma_{\mathrm{PTB}}$ presented in this work these are $\alpha_{\mathrm{MIT}}$ and $\gamma_{\mathrm{MIT}}$ from Ref.~\cite{Hur2022evidence}, the
$^{2}\!S_{1\!/\!2}\!\rightarrow\!{}^{2}\!D_{3\!/\!2}$ transition in Yb$^{+}$~\cite{Counts2020evidence} (denoted $\beta$), ${}^{1}\!S_{0}\!\rightarrow\!{}^{3}\!P_{0}$ in Yb~\cite{Ono2022observation} (denoted $\delta$), and ${}^{1}\!S_{0}\!\rightarrow\!{}^{1}\!D_{2}$ in Yb~\cite{Figueroa2022precision} (denoted $\epsilon$).
In total, we construct 7 mass-normalized isotope-shift vectors $\mIS{\boldsymbol{\nu}}_\tau$, with $\tau\in\{\alpha_{\mathrm{PTB}}, \alpha_{\mathrm{MIT}}, \beta, \gamma_{\mathrm{PTB}}, \gamma_{\mathrm{MIT}}, \delta, \epsilon\}$. 
Since these vectors are four-vectors, they are uniquely described by their projections onto four basis vectors. 
We choose the basis vectors $\mIS{\boldsymbol{\nu}}_\delta$ and $\boldsymbol{1}$, which span the plane of King linearity [see Eq.~\eqref{eq:kingrelation_vector}, with $\alpha\to \delta$], and  $\Lp$ and $\Lm$ (defined in the Supplemental Material), 
which are orthogonal to this plane~\cite{Hur2022evidence}.
We obtain $\mIS{\boldsymbol{\nu}}_\tau = 
        F_{\tau\delta} \mIS{\boldsymbol{\nu}}_\delta + K_{\tau\delta} \boldsymbol{1}  + \lambda_+^{(\tau)} \Lp + \lambda_-^{(\tau)} \Lm$
with $F_{\tau\delta}$ and $K_{\tau\delta}$ as given in Eq.~\eqref{eq:kingrelation_vector} and the coordinates $(\lambda_+^{(\tau)}, \lambda_-^{(\tau)})$ characterizing the deviation of the isotope shift $\boldsymbol{\nu}_\tau$ from the linear relation in Eq.~\eqref{eq:kingrelation_vector}.
As shown in Fig.~\ref{fig:NDP}\,\textbf{a}, the data points $(\lambda_+^{(\tau)}, \lambda_-^{(\tau)})$ lie to a good approximation on the solid black line through the origin of the $(\lambda_+, \lambda_-)$ plane. 
This implies that the tension of the Yb isotope-shift data with respect to King linearity can to a large extent be explained by a nonlinearity source (new physics, for example) with the appropriate slope $\lambda_-/\lambda_+$ in the decomposition plot. 
We compare the slope of the linear fit to the slopes predicted by the new physics term (dash-dotted line) and the quadratic field shift $\drtwo^2$ (dotted line, from experimental $\drtwo$ data~\cite{ANGELI201369}). 
Both have uncertainties not visible in Fig.~\ref{fig:NDP}\textbf{a}, so we conclude that neither can be the leading source of the nonlinearity in the Yb King plot.
Details of the nonlinearity decomposition are provided in the Supplemental Material.
Another candidate is nuclear deformation~\cite{Allehabi:2020xgf,Hur2022evidence}, in particular $\drfour$, which we predict using an \emph{ab initio} approach.

{\it Nuclear structure effects.--}
Recent developments have made heavy nuclei accessible to \textit{ab initio} nuclear structure calculations~\cite{Shimizu2021qvsm,Miyagi2022no2b3n}. 
To predict $\drfour$, we use the valence-space in-medium similarity renormalization group (VS-IMSRG)~\cite{Hergert2016imsrg,Stroberg2019vsimsrg} together with the quasi-particle vacua shell model (QVSM)~\cite{Shimizu2021qvsm} to solve the many-body Schr\"odinger equation.
We employ nucleon-nucleon and three-nucleon interactions from chiral effective field theory, using the so-called 1.8/2.0 (EM)~\cite{Hebeler2011magic} and $\Delta\text{N}^2\text{LO}_{\text{GO}}$~\cite{Jiang2020deltan2logo} interactions, which differ in their construction and how they are fit to data, to give insight into interaction uncertainties. 
To assess many-body uncertainties, we solve the VS-IMSRG at the two- and three-body truncations~\cite{Heinz2021imsrg3} and employ two different valence spaces, VS1 with a \elem{Sn}{132} core and VS2 with a \elem{Gd}{154} core, in the QVSM. 
In Fig.~\ref{fig:NDP}\,\textbf{a}, we show the prediction of the nonlinearity from our nuclear structure calculations, where the uncertainty is represented by the gray band. This uncertainty stems from a correlated statistical model accounting for interaction and many-body uncertainties including correlations between isotope pairs.
Details of our nuclear structure calculations and the uncertainty quantification are provided in the Supplemental Material.
The two sets of calculations using the 1.8/2.0~(EM) Hamiltonian with valence spaces VS1 and VS2 serve as representative samples of our nuclear theory predictions. 
Since the best-fit line is compatible with the \textit{ab initio} calculations for $\drfour$, we assume $\drfour$ to be the leading King-plot nonlinearity in Yb.  

Combining isotope-shift measurements, nuclear mass-ratio measurements, and charge radius measurements~\cite{ANGELI201369} with atomic structure calculations of $G_\alpha^{(4)}$ [see Eq.~\eqref{eq:isotopeshift_all}] using AMBiT~\cite{kahl19cpc}, we recast the Yb King-plot analysis into a measurement of nuclear deformation, which can be used to benchmark nuclear structure calculations. The procedure is given in the Supplemental Material, and the extracted changes in \drfour{} relative to the reference value of $\drfour^{176,174} = 7\,\fm^4$ (star) are shown in Fig.~\ref{fig:NDP}\,\textbf{b}. This reference value is based on input from both our \textit{ab initio} results and density functional theory calculations~\cite{Hur2022evidence} that all predict $\drfour^{176,174} = 6$--$8\,\fm^4$. The experimental data show that the evolution of $\drfour{}^{A,A-2}$ along the isotope chain is nearly flat, remarkably consistent with our \textit{ab initio} calculations within uncertainties.
Nonetheless, from the residuals for the transitions $\alpha_{\mathrm{PTB}}$, $\beta$, $\gamma_{\mathrm{PTB}}$, $\gamma_{\mathrm{MIT}}$ and $\epsilon$, shown in Fig.~\ref{fig:NDP}\,\textbf{a}, we deduce a $23\,\sigma$ preference for more than one linearity, leaving open the possibility of a new boson being responsible for the next-to-leading King nonlinearity. This strengthens the prior two-source hypothesis~\cite{Hur2022evidence} by a factor of more than 5.

{\it Bounds on new physics.--}
To extract bounds on the hypothetical new boson, we combine our isotope-shift measurements and nuclear mass measurements with the isotope-shift measurements of Ref.~\cite{Ono2022observation}.
This allows us to eliminate both the charge radius variance $\delta \langle r^2 \rangle$ and $\drfour$ from the system of isotope-shift equations without requiring theoretical input. Assuming higher-order SM terms beyond $\drfour$ to be negligible, the generalized King-plot~\cite{Berengut:2020itu} can be used to set an upper bound on the new physics coupling $\alpha_\mathrm{NP}$ as a function of the mass $m_\phi$ of the new boson. Note that King-plot bounds should always be understood in the context of a given dataset since they are highly sensitive, not only to the central values and uncertainties of the frequency and mass measurements, but also to the unknown nuclear and electronic effects that are reflected in the isotope shift data. These aspects are discussed in more detail in the Supplemental Material.

\begin{figure}[ht!]
    \centering
    \includegraphics[width=3.4in]{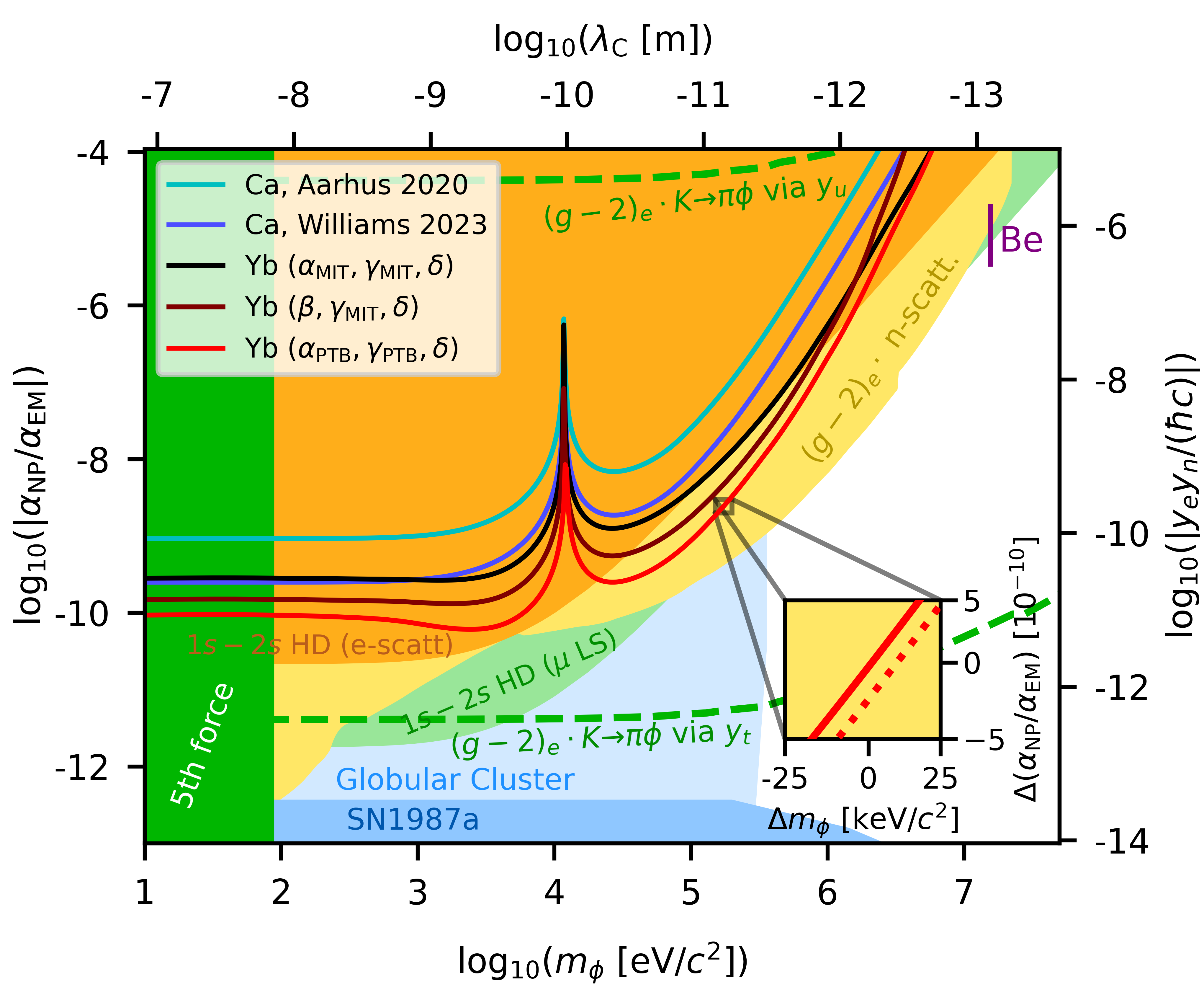}
    \caption{Exclusion plot for the new boson $\phi$ coupling to electrons and neutrons. Solid lines: $2\sigma$ upper bounds from Ca$^+$ King plots (cyan: Aarhus~\cite{Solaro2020improved}, blue: Williams~\cite{Chang:2023teh}), Yb($^+$) King plots (black, maroon:~\cite{Counts2020evidence,Hur2022evidence,Ono2022observation}), and from this work (red) on the product of couplings $y_ey_n/(\hbar c) = 4\pi \alpha_{\rm{NP}}$ to electrons and neutrons depending on the new boson's mass $m_\phi$. 
    The inset shows an extract of the $2\sigma$ upper bound (solid red line) and the $2\sigma$ lower bound (dotted red line) produced by this work.
    The shaded areas are disfavored by the constraint on $y_e$ from $(g-2)_e$~\cite{Hanneke:2008tm,Fan:2023gm2electronNorthwestern}, times the constraints on $y_n$ from neutron optics~\cite{Leeb:1992qf} and neutron scattering experiments~\cite{Nesvizhevsky:2007by,Pokotilovski:2006up,Barbieri:1975xy,Leeb:1992qf} (yellow), hydrogen and deuterium spectroscopy using electron scattering or the Lamb shift in muonic atoms to determine the charge radius~\cite{Frugiuele:2016rii} (see also Ref.~\cite{Potvliege:2023lvf}), fifth force searches~\cite{Bordag:2001qi,Bordag:2009zz}, stellar evolution in the globular cluster~\cite{Grifols:1986fc,Grifols:1988fv}, and energy loss in the supernova SN1987A~\cite{Raffelt:2012sp}. The dashed curves show the constraint on $y_e$ from $(g-2)_e$ times the constraint on $y_n$ from $K\to \pi + \mathrm{invisible}$, assuming the new boson $\phi$ couples only to the top-quark (via $y_t$) or only to the up-quark (via $y_u$). The purple band labeled ``Be'' indicates the coupling range required for a protophobic boson with $m_\phi=$ 17\,MeV/$c^2$ to explain the ATOMKI anomaly~\cite{Krasznahorkay2016_beryllium_anomaly}.
    }
    \label{fig:exclusion_plot}
\end{figure}

The new bound provided by this work is shown in Fig.~\ref{fig:exclusion_plot} in red.
If the second King-plot nonlinearity were to be explained by new physics only, the couplings of the new boson would be expected to reside between the $2\sigma$ upper bound (solid red line) and the $2\sigma$ lower bound (dotted red line), which are only distinguishable in the inset. 
The new bound supersedes, both in precision and in magnitude, the Ca$^{+}$ King plot bounds~\cite{Solaro2020improved,Chang:2023teh} and the generalized King plot bounds using previous isotope shift data in Yb$^+$~\cite{Counts2020evidence,Hur2022evidence,Ono2022observation} in combination with the new mass-ratio measurements provided by this work (see Table~\ref{tab:joint_table}). Moreover, it takes into account the significant shifts in the mass-ratio measurements $\eta_A$ and the isotope shifts $\nu_\alpha^{A,A+2}$ highlighted in Table~\ref{tab:joint_table}. For instance, the differences in the $\nu_\alpha^{A,A+2}$ measurements entering the bound from the dataset $(\alpha_\mathrm{MIT}, \gamma_\mathrm{MIT}, \delta)$ (shown in black) and the new bound (red) lead to the displacement of the characteristic peaks at the high end of the plotted $m_\phi$ values, resulting in different slopes. A detailed comparison of our new bound with respect to the bounds presented in Refs.~\cite{Counts2020evidence,Hur2022evidence}, alongside a discussion of the competing astrophysical and laboratory bounds, which are shown in Fig.~\ref{fig:exclusion_plot} as exclusion regions and green dashed curves, can be found in the Supplemental Material.

{\it Discussion and conclusion.--}
We measured the isotope shifts for the $^{2}\!S_{1\!/\!2}\!\rightarrow\!{}^{2}\!D_{5\!/\!2}$ electric quadrupole transition and the $^{2}\!S_{1\!/\!2}\!\rightarrow\!{}^{2}\!F_{7\!/\!2}$ electric octupole transition
on five stable, spinless $\mathrm{Yb}^+$ isotopes and with relative uncertainties on the order of $10^{-8}$ to $10^{-9}$, as well as mass ratios of these isotopes to a precision of $4\times 10^{-12}$.

Our measurements, combined with atomic structure calculations, have enabled a first direct extraction of the evolution of $\delta\langle r^4 \rangle$ across the ytterbium isotope chain. To understand whether the change in the nuclear charge distribution is consistent with the strong interaction, we have performed {\it ab initio} calculations based on chiral effective field theory interactions. Our results reproduce the experimental $\delta\langle r^4 \rangle$ remarkably well for such heavy nuclei.
The $\langle r^4 \rangle$ nuclear charge moment can provide information about both nuclear deformation~\cite{Allehabi:2020xgf} (because Yb nuclei are prolate deformed) and the surface thickness of the nuclear density~\cite{reinhard20prc,Kurasawa:2020fli}. It has also been shown that this charge moment can provide insights to experimentally estimate the neutron-skin thickness~\cite{Kurasawa:2020fli,naito21prc}.
An exciting future direction is to advance this work to the neutron-rich calcium isotopes. This can widen the search for the new boson and shed light on the puzzling increase of charge radii towards $^{52}$Ca~\cite{GarciaRuiz:2016ohj}.

Our isotope-shift and nuclear mass measurements, combined with the isotope-shift measurements of Ref.~\cite{Ono2022observation} allow us to set competitive spectroscopy bounds on new bosonic mediators between neutrons and electrons using the generalized King-plot method~\cite{Berengut:2020itu}. 
With increasing experimental precision, data from more isotopes will be needed to overcome atomic and nuclear structure uncertainties.
In ytterbium, a possible candidate is \elem{Yb}{166}, with a half-life of 54 hours. 
Another option is to use elements Sn~\cite{leibrandt2022prospects} or Xe~\cite{rehbehn23prl}, each with seven spinless, stable isotopes and suitable clock transitions that could be found in different ionization stages~\cite{lyu2023ultrastable}. %

{\it\textbf{Acknowledgments.--}}
We thank the referees for very valuable feedback on the manuscript. We thank Joonseok Hur, Martin Steinel and Vladan Vuleti\'{c} for helpful discussions.
This work is supported by the Max Planck Society (MPG), the International Max Planck Research Schools for Precision Tests of Fundamental Symmetries (IMPRS-PTFS), the European Research Council (ERC) under the European Union’s Horizon 2020 research and innovation programme under Grant Agreements 832848 (FunI) and 101020842 (EUSTRONG), and by the Deutsche Forschungsgemeinschaft (DFG, German Research Foundation) – Project-ID 273811115 – SFB 1225 ISOQUANT as well as under Germany’s Excellence Strategy – EXC-2123 QuantumFrontiers – 390837967 (RU B06) and through Grant No.~CRC 1227 (DQ-\textit{mat}, projects B02 and  B03). This work has been supported by the Max-Planck-RIKEN-PTB-Center for Time, Constants, and Fundamental Symmetries. L.S.D.\ acknowledges support from the Alexander von Humboldt Foundation. This work comprises parts of the Ph.D.~thesis work of M.D.\ submitted to Heidelberg University, Germany, parts of the Ph.D.~thesis work of C.-H.Y.\ submitted to Leibniz Universität Hannover, Germany, and parts of the Ph.D.~thesis work of M.H.\ submitted to the TU Darmstadt, Germany. 
M.H., T.M., and A.S.\ gratefully acknowledge the computing time provided on the high-performance computer Lichtenberg at the NHR Centers NHR4CES at TU Darmstadt. 
N.S. acknowledges the support of the ``Program for promoting research on the supercomputer Fugaku'', MEXT, Japan (JPMXP1020230411), and the MCRP program of the Center for Computational Sciences, University of Tsukuba (NUCLSM). 
E.F.\ and F.K.\ thank CERN for their hospitality during the early phase of this work.

\bibliographystyle{apsrev4-1}
\bibliography{references.bib}

\clearpage

\appendix

\begin{center}
    {\scshape {\Large Appendix}}
\end{center}

\tableofcontents
\section{Experimental setup for the mass-ratio determinations}\label{appen:mass_exp_setup}

The experimental setup of \penta as well as the measurement schemes and techniques are similar to those in prior measurements. Details can be found in~\cite{repp2012_pentatrapscheme, roux2012_pentatower, schussler2020_meta, filianin2021_reQ, kromer2022_Pbmass, kromer2024_238Umass} and just a short overview and details of this measurement are given.

The mass-ratio determinations to the precision presented rely on the use of highly charged ions to increase the cyclotron frequency. The highly charged $^{A}\textrm{Yb}^{42+}$ ions are produced in a room-temperature electron beam ion trap via in-trap laser ablation~\cite{schweiger2019_tipebit}, allowing isotope selection via multi-position targets with isotope-enriched samples of YbO. A cloud of highly charged ions is extracted with a kinetic energy of approximately $4$\,keV/$q$ and transported through an electrostatic beamline equipped with a Bradbury-Nielsen Gate~\cite{bradbury1936_bng_nielsen_gate} for time-of-flight selection of a single charge state. The initial kinetic energy is reduced to a few eV/$q$ via two pulsed drift tubes, one above the magnet and one right above the Penning-trap assembly inside the magnet. At this energy, the ions can be caught in the upper trap using a low-voltage potential-lift scheme.

\begin{figure*}[tbh!]
    \includegraphics[width=0.99\linewidth]{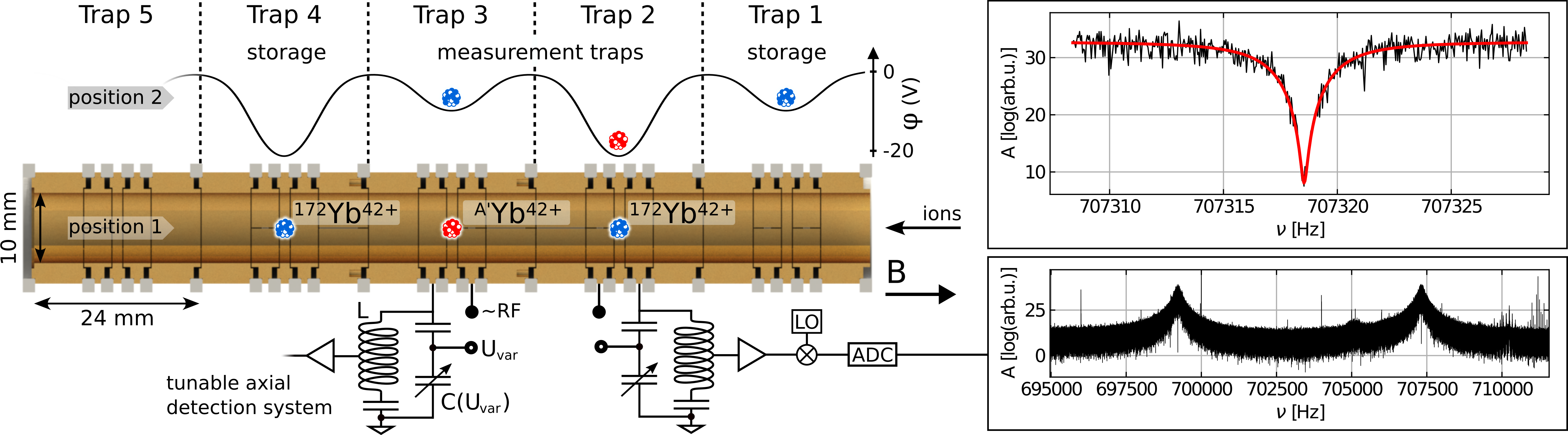}
    \caption{Overview of the \penta measurement setup with a sectional view of the trap tower, approximate schematics of the detection system, and example spectra of the axial detection in trap 2 for the determination of the cyclotron frequency ratio of $^{176}\textrm{Yb}^{42+}$ over $^{172}\textrm{Yb}^{42+}$. The lower spectrum shows the detection system resonance frequency tuned to match one of the two axial frequencies of the respective ion during the measurement in position 1 and position 2. The upper spectrum shows a zoom-in of one axial dip spectrum of $^{172}\textrm{Yb}^{42+}$ with the model fit given as a red line.}
    \label{fig:pentasetup}
\end{figure*}

The cryogenic setup with a stack of five identical Penning traps is placed in a superconducting magnet with a field strength of $7\,\text{T}$. The traps and detection electronics, see Fig.~\ref{fig:pentasetup}, are cooled to the temperature of liquid helium ($4$\,K) providing a low noise environment for ion detection and cooling. The ions are loaded with deliberate low transport efficiency to load only a single ion per shot into the trap tower. 

For maximal relative precision (highest possible cyclotron frequency), we used one of the highest charge states of 42+ produced by our ion source. Given the field strength of the superconducting magnet and the resonance frequencies of the detection systems, the eigenfrequencies of trapped $^{A}\textrm{Yb}^{42+}$ ions are approximately $26$\,MHz for the trap-modified cyclotron frequency, $700$\,kHz and $475$\,kHz for the axial frequencies, and $9.5$\,kHz and $4.3$\,kHz for the magnetron frequencies for traps 2 and 3, respectively. A summary of all trapping parameters is given in Tab.~\ref{tab:trapparams}.

\begin{table}[tb]
\caption{Trap parameters relevant in the analysis of the systematic effects. The approximate eigenfrequencies are given for the common reference ion $^{172}\textrm{Yb}^{42+}$. The magnetic field was calculated using the cyclotron frequency $\nu_c$ and ion's charge-to-mass ratio from literature~\cite{huang2021_ame}.}
\begin{ruledtabular}
\begin{tabular}{lll}
Parameter &\textrm{Trap 2}&\textrm{Trap 3} \\ \hline  \\ [-1.7ex]
$r_0$\,[mm]~\cite{roux2012_pentatower}&5.000\,(5)&5.000\,(5)\\
\textit{TR}\,[1]&0.87966\,(15)&0.879002\,(7)\\
$U_0$\,[V]&$-28.14$&$-12.85$\\
$T_z$\,[K]& 5\,(2) & 8\,(2)\\
$\rho_{+,exc}$\,[{\textmu}m]& 12\,(2)& 12\,(2)\\
RLC $Q$-factor\,[1]&$\approx$\,3300&$\approx$\,9800\\
$\nu_+$\,[MHz]&$\approx$\,26.26&$\approx$\,26.27\\
$\nu_z$\,[kHz]&$\approx$\,707.3&$\approx$\,478.1\\
$\nu_-$\,[kHz]&$\approx$\,9.5&$\approx$\,4.4\\
$B_0$\,[T]&7.002\,15\,(2)&7.002\,16\,(2)\\
$B_2$\,$\left[\frac{\text{mT}}{\text{m}^2}\right]$&28\,(2)&$-5\,(2)$\\
$c_2$\,$\left[{10^{-3}\,\text{mm}^{-2}}\right]$&$-14.88576\,(1)$&$-14.89708\,(1)$\\
$c_4/c_2$\,$\left[{10^{-5}\,\text{mm}^{-2}}\right]$&$-3\,(8)$&$-7\,(8)$\\
$c_6/c_2$\,$\left[{10^{-6}\,\text{mm}^{-4}}\right]$&$-4\,(6)$&$-2\,(6)$\\
\end{tabular}\label{tab:trapparams} 
\end{ruledtabular}
\end{table}

The axial eigenmotion is resistively cooled via the detection system: The induced image current by an ion creates a voltage drop across the high impedance of the connected superconducting resonance circuit (LCR). This voltage signal acts as negative feedback dampening the ion's axial motion, limited by the thermal noise of the detection systems at approximately $4.2$\,K. Radial modes are indirectly cooled by energy exchange to the axial mode by sideband coupling through a quadrupolar RF field~\cite{Cornell1990_pnp_coupling}.

The supposedly single ions are cleaned from possible contaminant ions by using broadband magnetron excitation with simultaneous magnetron sideband cooling of the ion of interest and trap potential variation. Following this procedure, a set of three ions is prepared in an $A$-$A'$-$A$ sequence in the three central traps, and these are reloaded on average every second day due to charge exchange with rest gas atoms.

For the measurement of the axial frequencies, the dip method is applied~\cite{wineland1975_dipfit}: When the ion reaches thermal equlibrium with the detection circuit, the trapped ion acts equivalent to a parallel connected series LC-cirucit with a low impedence at the axial frequency. With the ions axial frequency matched to the resonance frequency of the detection circuit, a dip in the resonance can be fitted to extract the axial frequency, see Fig.~\ref{fig:pentasetup}. Because of the difference in charge-to-mass ratio, the axial frequencies of different isotopes in the same charge state, stored at the same trapping potential, are separated by multiple kHz. This difference is bridged by tuning the resonance frequency of the LCR circuit using varactor diodes (voltage controlled capacitances)~\cite{heisse2017_protonmass_varactor}, see Fig.\,\ref{fig:pentasetup}. In this way, the trap potential settings can be kept identical for two ions with a charge-to-mass ratio difference of up to 3\%, ensuring identical field imperfections for both ions and suppressed systematic shifts on the determined cyclotron frequency ratio. 

For the radial modes the double-dip method is used and for the precise determination of the trap-modified cyclotron frequency, the phase-sensitive Rabi-type Pulse and Phase (PnP) technique~\cite{Cornell1989_pnp_nitrogen} is applied. The initial phase imprint for the PnP method is realized by dipolar excitation of the trap-modified cyclotron mode, increasing its amplitude $\rho_+$ on the order of $12$\,µm. This is followed by a free evolution of the phase and finally coupling of the radial and axial modes to measure the evolved phase using the axial detection system. A reference phase at a short phase evolution time is subtracted from the long evolution time phase to cancel phase offsets in the measurement chain. The equality of excitation radii for the different ions has been tested to a level of 2\%, which ensures a low impact of systematic shifts dependent on the excitation radii on the final cyclotron frequency ratio. During the long phase evolution time, the dip spectrum is recorded, resulting in a simultaneous determination of the two higher eigenfrequencies. The ion species is switched inside both measurement traps by transporting the three ions up or down via potential variation, with one ion always in one of the two storage traps 1 or 4. 

Traps 2 and 3 are used simultaneously as measurement traps which increases statistics and allows for cross-checks of trap and detection system related systematic shifts. The determination of the cyclotron frequencies of two ions is done sequentially. To switch ion species in the measurement traps the set of ions is adiabatically transported to the neighboring traps, see positions 1 and 2 in Fig.~\ref{fig:pentasetup}.

At the beginning of each measurement cycle, all eigenfrequencies are determined roughly via dip and double-dip methods for both types of ions to determine measurement settings for the PnP method and tune the detection system on resonance with the axial frequencies. The automated measurement cycle is split into two parts: The trap-modified cyclotron frequency is determined via multiple PnP-type phase measurements with a set of up to 10 different evolution times between $0.1$ and $100$\,s, each phase measurement repeated for 4 to 6 times, allowing one to unwrap the absolute frequency. In the second part, typically running for 12 hours before repeating the absolute frequency determination, only the shortest and longest of these evolution times are measured to follow the change of the trap-modified cyclotron frequency due to changes in the magnetic field. This is possible due to our liquid helium level and pressure-stabilized superconducting magnet, resulting in a shot-to-shot $\nu_+$ frequency stability on the level of $3$--$5 \times 10^{-11}$. During these 12 hours, the ion species is switched between 20 and 30 times in each trap. 

The cyclotron frequency $\nu_{c}=qB/(2\pi m)$ of an ytterbium ion with mass $m$ and charge $q$ in a magnetic field $B$ is determined from all measured eigenfrequencies with $\nu_c^2 = \nu_+^2 + \nu_z^2 + \nu_-^2$~\cite{brown_gabriels1982_invariance}, with the magnetron frequency $\nu_-$ determined in the preparation of the measurement cycle with sufficient precision due to the frequency hierarchy and respective error propagation. Every subsequent cyclotron frequency measurement of the two ions yields a cyclotron frequency ratio. The ratio is calculated from interpolated values of the cyclotron frequency measurement of one ion species to the time-stamp of the other ion species to adjust to first-order magnetic field drifts.

With the measured $\mathrm{R}_A^{\text{CF}}$ the nuclear mass ratios are derived via
\begin{equation}	
\eta_A = \mathrm{R}_A^{\text{CF}} + \frac{(\mathrm{R}_A^{\text{CF}}- 1) \left[28m_e - E_{172}^{(28)} /c^2\right] - \Delta E_A^{(28)} /c^2}{m(^{172}\text{Yb})-70m_e+E_{172}^{(70)}/c^2}\ . \label{eq:mass_nuclear_ratio}
\end{equation}
Here, $m(^{172}\text{Yb})$~\cite{huang2021_ame} is the mass of the neutral atom and $m_e$~\cite{tiesinga2021_codata_emass} the electron mass. $E_{172}^{(28)}$ and $E_{172}^{(70)}$ are the binding energies of the 28 electrons in the $^{172}\text{Yb}^{42+}$ ion and 70 electrons in the $^{172}\text{Yb}$ atom, respectively. Thus, the denominator in Eq.~\eqref{eq:mass_nuclear_ratio} represents the nuclear mass of $^{172}$Yb, i.e., $m(^{172}\text{Yb}^{70+})$, or $m_{172}$ in the main text. The necessary electron binding energies and their dependence on isotopes $\Delta E_A^{(28)}\!=\!E_{172}^{(28)}\!-\!E_{A}^{(28)}$ are calculated with the GRASP2018 code~\cite{GRASP2018} discussed in later sections.

\begin{table}[tb]
	\centering
	\caption{Neutral masses in atomic mass units of the five even ytterbium isotopes determined from the measured cyclotron frequency ratios $\text{R}^{\text{CF}}_A = \nu_{c,172}/\nu_{c,A}$, the binding energy to of the 42 missing electrons $E_{172}^{(70)} - E_{172}^{(28)}$, and the literature mass of isotope $m(^{172}\text{Yb})$~\cite{huang2021_ame}. For comparison, we also provide the literature mass values listed in the atomic mass evaluation 2020~\cite{huang2021_ame}.}
 \begin{ruledtabular}
	\begin{tabular}{lll}
		Isotope $A$ & $m(A)$ (this work) & $m(A)$ \cite{huang2021_ame} \\ \hline \\ [-1.7ex]
		168 & 167.933\,890\,939\,(14) & 167.933\,891\,297\,(100) \\ 
		170 & 169.934\,767\,218\,(14) & 169.934\,767\,242\,\phantom{0}(11)\\ 
		174 & 173.938\,867\,541\,(14) & 173.938\,867\,545\,\phantom{0}(11)\\ 
		176 & 175.942\,574\,697\,(14) & 175.942\,574\,706\,\phantom{0}(15) \\ 
	\end{tabular}%
 \end{ruledtabular}
	\label{tab:neutralmasses}
\end{table}

While the mass ratio can be used in full precision for the King-plot analysis, the neutral masses of the isotopes can also be calculated using the presented binding energies. The resulting uncertainty is limited by the literature value for $m(^{172}\text{Yb})$ with a relative uncertainty of $8 \times 10^{-11}$. All neutral masses are given in Tab.~\ref{tab:neutralmasses} and are compared to the current literature values of the atomic mass evaluation 2020~\cite{huang2021_ame}. The masses of three isotopes agree within $1.5\sigma$ and the mass of $^{168}\text{Yb}$ shows a deviation by $3.5\sigma$. The uncertainty of $m(^{168}\text{Yb})$ is improved compared to the literature value by a factor of $7$.

\section{Systematic effects and analysis of the mass-ratio measurement data}\label{appen:mass_sys_ana}

A summary of all systematic shifts on the determined ratios can be found in Tab.~\ref{tab:pentasys} for both traps and the different mass ratios. Some of the given shifts are commonly known shifts and uncertainties on the real motional frequencies. Among these are the relativistic shift~\cite{Brown1986_geonium}, shifts due to field imperfections~\cite{Brown1986_geonium,ketter2014_fieldimperfections}, and the image charge shift~\cite{vandyck1989_ics, schuh2019_ics}. The trap parameters necessary to calculate these and other effects are given in Tab.~\ref{tab:trapparams}. The dipole excitation necessary for the phase imprint on the trap-modified cyclotron motion for the PnP technique was compared between two isotopes to check for equally increased amplitudes $\rho_+$ of this radial mode and yielded a maximum deviation of 2\%. Although the absolute excitation radius cannot be determined with such precision, the precision of the excitation radius ratio allows for a significant reduction in systematic uncertainties.

The uncertainty of the image charge shift effect originates from the most precise test to date of the respective analytical estimation of this effect, which agreed to a level of 5\% limited by the experimental uncertainty~\cite{schuh2019_ics}.
In addition to these, there are systematic uncertainties mainly originating from our detection system and measurement methods. The highest uncertainty originates in the determination of the axial frequency via the dip determination, which is sensitive to the precision with which the resonance circuit can be tuned to the axial frequency~\cite{rau2020_deuteron_mass_lineshape}. The determination of the resonance frequency is precise to only 2\,Hz in our setup. The given uncertainties on the determined ratio due to this effect are extracted from measurements of the axial frequency in dependence on the resonance frequency as well as full ratio determinations with deliberate detuning of the RLC resonance by 10--20\,Hz for only one ion. 

The nonlinear phase readout originates from a nonlinear transfer function of the ions cyclotron phase at the time of the PnP coupling pulse to the axial mode. The residual of the ideally linear transfer function follows a sinusoidal behavior with amplitudes of 0.25 and 0.5 degrees in trap 2 and 3, respectively. These phase offsets average out for the long accumulation time measurement, but create constant offsets for the short reference phase, which results in the given uncertainties.

The magnetron frequency systematic uncertainty originates from the calculation of the magnetron frequencies of one ion from the measured magnetron frequency of the other, using the literature mass values. This is sufficient, since the cyclotron frequency ratio is more sensitive to the magnetron frequency difference than to the absolute value, which is only determined to low precision before each measurement.

\begin{table}[tb]
    \centering
    \caption{Systematic corrections $\Delta \text{R}$ for the determined ratio $\text{R}^{\text{CF}} = \text{R}' + \Delta \text{R}$ with $\text{R}' = \nu_{c,172}/\nu_{c,A}$ being the measured ratio. All values are expressed in parts per trillion ($1 \cdot 10^{-12}$). For more details see text.}
    \begin{ruledtabular}
    \begin{tabular}{llrr}
    Effect & A & \multicolumn{1}{c}{Trap 2} & \multicolumn{1}{c}{Trap 3}  \\ \hline \\ [-1.7ex]
    Field imperfect ($B_i$, $c_i$) & all & $0.0\,(0.5)$ & $0.0\,(0.5)$ \\
    Nonlinear phase readout & all & $0.0\,(1.0)$ & $0.0\,(1.5)$ \\
    Magnetron frequency & all & $0.0\,(1.0)$ & $0.0\,(1.0)$ \\
    Dip lineshape & all & $0.0\,(4.0)$ & $0.0\,(5.0)$ \\ \hline & \\[-1.7ex]
    Common total & all & $0.0\,(4.4)$ & $0.0\,(5.3)$ \\ \hline & \\[-1.7ex]
    \multirow{4}{*}{Image charge shift} & 176 & \multicolumn{2}{c}{$-10.0\,(0.5)$} \\
     & 174 & \multicolumn{2}{c}{$-\phantom{1}4.9\,(0.2)$} \\
      & 170 & \multicolumn{2}{c}{$\phantom{-1}4.8\,(0.2)$} \\
       & 168 & \multicolumn{2}{c}{$\phantom{-1}9.5\,(0.5)$} \\ \hline & \\[-1.7ex]
    \multirow{4}{*}{Special relativity} & 176 & \multicolumn{2}{c}{$\phantom{-}1.0\,(1.5)$} \\
     & 174 & \multicolumn{2}{c}{$\phantom{-}0.5\,(1.5)$} \\
      & 170 & \multicolumn{2}{c}{$-0.5\,(1.5)$} \\
       & 168 & \multicolumn{2}{c}{$-1.0\,(1.5)$} \\
    \end{tabular}
    \end{ruledtabular}
    \label{tab:pentasys}
\end{table}

All systematic shifts are corrected in the weighted mean cyclotron frequency ratio results of the individual traps. None of the dominant systematic shifts and uncertainties are correlated between the two measurement traps. The final value is determined from the weighted mean of the results of the two traps.

\section{\textit{Ab initio} binding energy calculations for mass-ratio correction}\label{appen:mass_binding}

The electron binding energies of neutral Yb and Ni-like Yb$^{42+}$ can be obtained by summing the ionization potentials (IPs) of all the charge states listed in the NIST atomic database~\cite {kramida2021_nist_binding}. This results in the values 382,457\,(614) and 350,722\,(233)\,eV for the binding energies of the 70 and 28 electrons, respectively. To reduce uncertainties in these values, advanced atomic structure calculations were performed in this work. The procedure is straightforward for the case of highly charged Yb$^{42+}$. However, for neutral Yb, many close-lying levels render it difficult to treat the electron correlations accurately. Nevertheless, considering that the IPs of Yb and Yb$^{+}$ are experimentally known to be 6.254160\,(12)\,eV and 12.179185\,(25)\,eV~\cite{kramida2021_nist_binding}, respectively, we can partially avoid this problem by calculating the binding energy of Yb$^{2+}$. 

The calculations were performed via the \textit{ab initio} fully relativistic multiconfiguration Dirac-Hartree-Fock (MCDHF) and relativistic configuration interaction (RCI) methods~\cite{Grant1970,Desclaux1971,grant2007relativistic} implemented in the GRASP2018 code~\cite{GRASP2018,Jonsson2023-2}. The many-electron atomic state function (ASF)
is constructed as a linear combination of configuration state functions (CSFs) with common total angular momentum ($J$), magnetic ($M$), and parity ($P$) quantum \mbox{numbers:} $|\Gamma P J M\rangle = \sum_{k} c_k |\gamma_k P J M\rangle$.
Each CSF $|\gamma_k P J M\rangle$ is built from products of one-electron orbitals (Slater
determinants), $jj$-coupled to the appropriate angular symmetry and parity, and $\gamma_k$ represents orbital occupations, together with orbital and intermediate
quantum numbers necessary to uniquely define the CSF. $\Gamma$ collectively denotes \mbox{all} the $\gamma_k$ involved in the representation of the ASF. $c_k$ is the corresponding mixing coefficient. We first solve the MCDHF equations self-consistently~\cite{Grant1970,Desclaux1971,grant2007relativistic} under the Dirac-Coulomb Hamiltonian to obtain an ASF, represented by the set of $c_k$, together with the set of radial orbitals. Then the RCI method is employed to calculate the contributions from mass shift (or nuclear recoil effect), Breit interaction, frequency-dependent transverse photon interaction, and QED effects. 

The ground state of $^{172}$Yb$^{42+}$ is described by the configuration [Ar]$3d^{1\!0}~{}^1\!S_0$. In the single-configuration Dirac-Hartree-Fock calculation, one obtains a binding energy of 351,379.33\,eV for a point-like nuclear charge. The corresponding mass shift equals $-0.87\,(5)$\,eV with the uncertainty of the order of $(m_e/M)(\alpha Z)^4m_ec^2$ ($M$ is the mass of the nucleus). The Fermi model for nuclear-charge distribution gives rise to a correction of $-45.06\,(8)$\,eV due to the finite nuclear size effect, with the uncertainty resulting from the nuclear radius predictions. 
The Breit interaction further contributes energy of $-367.66$\,eV, and the frequency-dependent transverse-photon interaction adds $6.65$\,eV to the total binding energy. The residual Coulomb-Breit interaction, the so-called correlation energy, is significant as well but more complicated to account for. Thus, its contribution and uncertainty are discussed in detail in later paragraphs. Similar calculations were also performed for neutral $^{172}$Yb, with the corresponding values being summarized in Tab.~\ref{tab.mcdhf}. 

\begin{table*}[tb]
    \centering
    \caption{Different contributions to the total binding energies of $^{172}$Yb$^{42+}$ and $^{172}$Yb$^{2+}$: DHF$_0$, the DHF energy assuming a point-like nuclear charge; MS, the mass shift; FNS, the finite nuclear size effect; Breit, the frequency-independent transverse photon interaction; ${\omega}$TP, the frequency-dependent transverse photon interaction; QED, the QED contribution based on screened-hydrogenic model; SDc, the correlation energies arising from single and double electron substitutions; and HOc, the systematic effect from all unaccounted correlation effects. The values of the DHF$_0$, Breit, ${\omega}$TP, and SDc terms are basis dependent. After taking into account all correlation effects, the basis dependency will be resolved. Thus, their uncertainties are accounted for as a whole in the high-order correlation energy (HOc). The total binding energies are rounded up to integer numbers. All entries are shown in units of eV. }
    \begin{ruledtabular}
    \begin{tabular}{llccccccrcl}
       Ion & Ground state &  DHF$_0$ & MS & FNS   &   Breit & ${\omega}$TP & QED & SDc  & HOc & Total \\
       \hline
       \\[-1.7ex]
        $^{172}$Yb$^{42+}$ & [Ar]$3d^{10}~{}^1S_0$   & $351379.33$ &  $-0.87\,(3)$  & $-45.06\,(8)$ &   $-367.66$ & $6.65$ &   $249.4\,(3.5)$ & $48.76\,(1)$ &  $1.4\,(1.4)$ &  $350773\,(5)$ \\
      $^{172}$Yb$^{2+}$ & [Xe]$4f^{14}~{}^1S_0$  &   $382828.68$   &  $-0.90\,(3)$ & $-45.31\,(8)$  &   $-384.95$ & $6.79$ & $251.1\,(3.5)$ & $117.26\,(9)$ & $12\,(12)$ &  $382283\,(16)$
    \label{tab.mcdhf}
    \end{tabular}
    \end{ruledtabular}
\end{table*}

Furthermore, the QED effects are implemented in the GRASP2018 package via a screened-hydrogen-like model for vacuum polarization (VP) and self-energy (SE) contributions, respectively~\cite{grant2007relativistic}. With values of $-54.61$\,eV and $307.45$\,eV for the VP and SE corrections, respectively, they reduce the binding energy by $252.84$\,eV for the case of $^{172}$Yb$^{42+}$. By comparing these values with respect to the results obtained from a hydrogen-like model without screening, we find that the many-electron QED effects contribute of the order of $-15.65$\,eV. The uncertainty of these QED calculations can be inferred from \textit{ab initio} QED calculations in Be-like Xe$^{50+}$ and U$^{88+}$ with a sub-eV uncertainty~\cite{PhysRevA.90.062517,PhysRevLett.126.183001}. We find that the results obtained by the GRASP2018 code equal $99.03$\,eV and $625.02$\,eV, respectively, for the two Be-like ions, and they are larger by $0.85$\,eV and $8.61$\,eV, respectively, compared to the \textit{ab initio} QED results~\cite{PhysRevA.90.062517,PhysRevLett.126.183001}. Assuming a similar systematic error, we obtain a QED contribution of $-249.4\,(3.5)$\,eV to the total binding energy of $^{172}$Yb$^{42+}$. However, since QED effects are significant mainly for inner-shell electrons, a similar QED correction of $-251.1\,(3.5)$\,eV is obtained for the case of $^{176}$Yb$^{2+}$. 

To derive the electron correlation energy, we systematically expand the size of the CSF basis set by allowing single and double (SD) substitutions of electrons from the occupied orbitals of the ground-state configuration to the systematically increasing set of correlation orbitals. These correlation orbitals are added and optimized layer-by-layer~\cite{Jonsson2023} up to $n=10$ ($n$ is the principle quantum number), with the highest orbital angular momentum in each layer equals $n-1$. 
It is found that the increment of the correlation energy at each layer decreases exponentially as a function of $n$~\cite{lyu2023extreme}. This allows us to extrapolate the electron correlation energy to $n=\infty$. With this scheme, we obtain an SD correlation energy of $48.76\,(1)$\,eV and $117.26\,(9)$\,eV for $^{172}$Yb$^{42+}$ and $^{172}$Yb$^{2+}$, respectively. The uncertainties are derived from extrapolations based on different numbers of data points.

Nevertheless, the full binding energy contains correlations from electron exchanges beyond the SD substitution scheme, as well as from basis functions outside the model space. These contributions are difficult to evaluate but can be estimated from the ionization potentials (IPs) of low-charged ions. We found that, based on SD substitutions from the $4s$ orbital, the calculated IPs for Yb, Yb$^+$, and Yb$^{2+}$ are 0.80\,eV, 0.40\,eV, and 0.65\,eV smaller than the experimental values (we note that the SD substitutions from orbitals below the $4s$ orbital generate more than 5 million CSFs for Yb$^+$ and Yb$^{3+}$, which is beyond the capacity of
available computing power at our disposal). As a conservative estimation (taken here as the difference between
calculated and experimental values, as discussed above), one can assume that the average shift of the calculated IPs for ions from Yb$^{2+}$ to Yb$^{23+}$ will not be larger than 0.8\,eV. To cover this effect, we add a shift of $8.8\,(8.8)$\,eV to the correlation-energy difference between Yb$^{2+}$ and Yb$^{24+}$. Since this shift is obtained based on the SD substitution from the $4s$ orbital, for the total-binding-energy calculations based on SD substitution from the $1s$ orbital, one has to add another $1.0\,(1.0)$\,eV shift to account for the contributions from core-core and core-valence correlations. Furthermore, the average shifts due to electron correlations in the theoretical IPs for ions between Yb$^{24+}$ and Yb$^{41+}$ can be constrained based on the calculations of Pd-like Xe$^{8+}$ and Cu-like Kr$^{7+}$, respectively. Since the QED effects are negligible for the IPs of Xe$^{8+}$ and Kr$^{7+}$, the discrepancy between experimental and calculated values arises from the unaccounted correlation effects. With SD substitutions from the $1s$ orbital, we arrive at an upper limit of 0.1\,eV~\cite{kromer2022_Pbmass} for the average shift in IPs. Thus, a shift of $0.9\,(9)$\,eV has to be added to the correlation-energy difference between Yb$^{24+}$ and Yb$^{42+}$. To be conservative, we also assume an average of 0.1 eV correlation-energy shift in the IPs for ions with charge states higher than Yb$^{42+}$. In total, the unaccounted high-order correlation effects are derived to be 1.4\,(1.4)\,eV and 12\,(12)\,eV for Yb$^{42+}$ and Yb$^{2+}$, respectively. 

Therefore, electron correlation effects contribute $50.2\,(1.4)$\,eV and $129\,(12)$\,eV to the total binding energy of Yb$^{42+}$ and Yb$^{2+}$, respectively. Altogether, we obtain $E_{172}^{(28)}=350{,}773\,(5)$\,eV and $E_{172}^{(68)}=382{,}283\,(16)$\,eV, where we have rounded up the numbers to integer values of eV. Adding up the IPs of the outermost two electrons, one arrives at $E_{172}^{(70)}=382{,}301\,(16)$\,eV for the total binding energy of the neutral atom. These results agree with the values obtained from the NIST atomic database but with the uncertainties reduced by more than 2 orders of magnitude. 

Since each Yb isotope bears a different mass and nuclear size, its total binding energy entails isotope shifts arising from different nuclear recoil and field effects. To investigate the isotope dependence of $\Delta E_A^{(28)}$, see Eq.~\ref{eq:mass_nuclear_ratio}, we start with a Dirac-Hartree-Fock (DHF) and RCI calculation where only the ground state configuration is considered for each isotope and charge state. With $^{172}$Yb$^{42+}$ as a reference, we obtain $\Delta E_A^{(28)}=-0.50\,(1)$\,eV, $-0.25\,(1)$\,eV, 0.25\,(1)\,eV, and 0.50\,(1)\,eV for isotopes $A=168$, $170$, $174$ and $176$, respectively. 

\section{Experimental details for frequency measurements}\label{appen:freq_exp_setup}

We determined the isotope shifts from absolute frequencies $\nu^{A}_{\alpha, \gamma}$ of the $^{2}\!S_{1\!/\!2}\!\rightarrow\!{}^{2}\!D_{5\!/\!2}$ electric quadrupole ($\alpha$) and the $^{2}\!S_{1\!/\!2}\!\rightarrow\!{}^{2}\!F_{7\!/\!2}$ electric octupole ($\gamma$) transitions in the stable even isotopes of $^{A}$Yb$^{+}$ ($A \in \{168,170,172,174,176\}$). 
For this, we measured the optical frequency ratios $\mathcal{R}_{\alpha, \gamma} = \nu^{A}_{\alpha, \gamma}/\nu^{171}_{\gamma}$ with $\nu^{171}_{\gamma}$ corresponding to the frequency of the $\gamma$ transition between the $F=0$ and $F=3$ hyperfine states of the $^{171}$Yb$^{+}$ isotope~\cite{Lange2021improved}. We determined the absolute frequencies by multiplying $\mathcal{R}$ to the recommended frequency value $\nu^{171}_{\gamma}= 642{,}121{,}496{,}772{,}645.12\,(12)$\,Hz~\cite{BIPM2022} and correcting for systematic frequency shifts.

\begin{figure}[tbh!]
    \centering
    \includegraphics[width=0.9\linewidth]{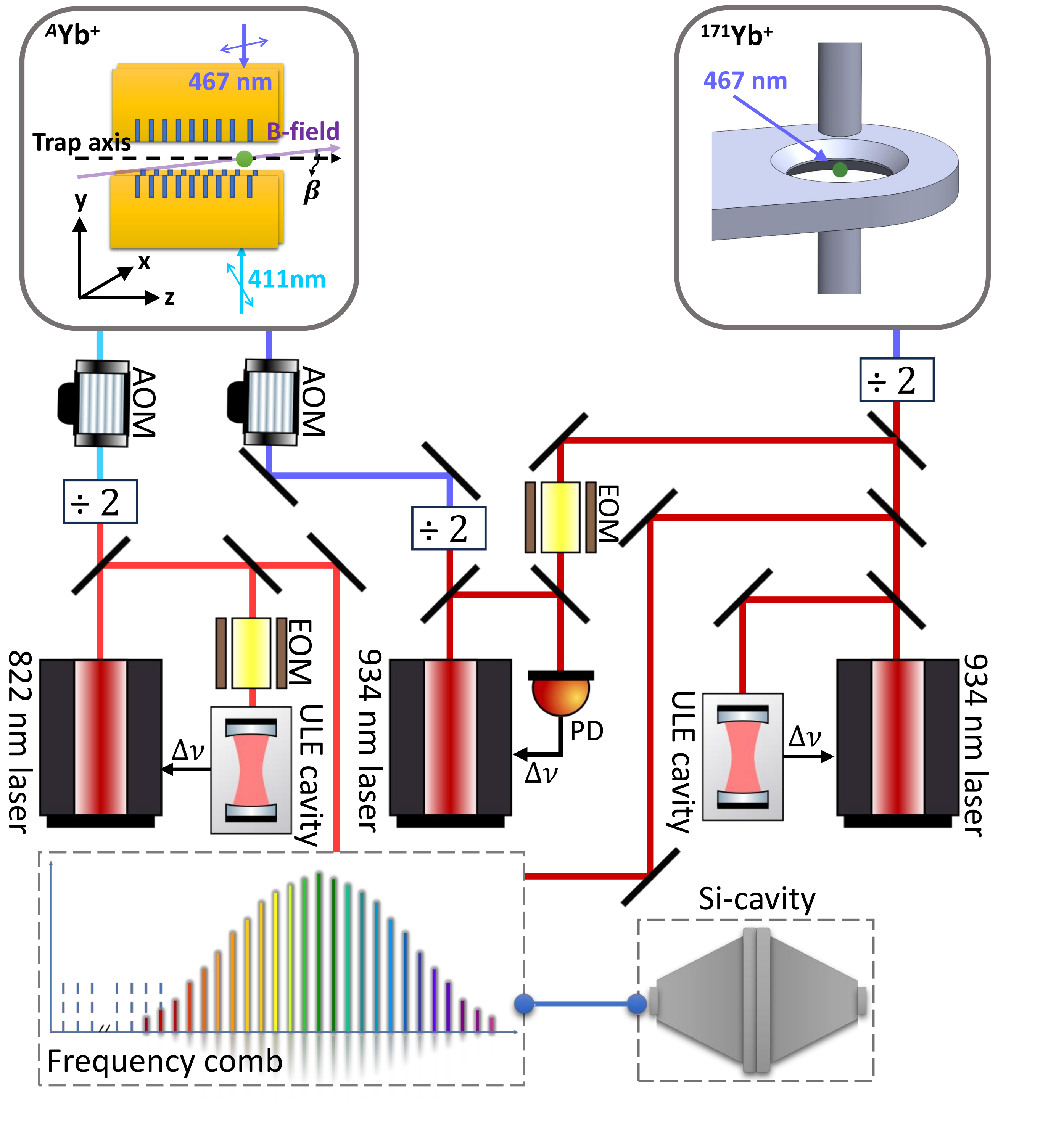}
    \caption{Simplified schematic representation of the experimental setup used for the interrogation of the $\alpha$ and the $\gamma$ transitions near 411\,nm and 467\,nm, respectively. Even isotopes of Yb$^{+}$ are trapped in a segmented linear radio-frequency (rf) Paul trap~\cite{pyka_high-precision_2014,keller_probing_2019} and the odd isotope $^{171}$Yb$^{+}$ is trapped in a ring style rf Paul trap~\cite{beaty1987simple}. Probe lasers in the infrared are referenced to ultra-low expansion (ULE) optical cavities for short-term stability, then further stabilized to a cryogenic silicon cavity~\cite{Matei2017}. Electro-optic modulators (EOMs) are used to bridge the isotope frequency differences and acousto-optic modulators (AOMs) are used to fine-tune the laser frequencies. A frequency comb setup is used to perform the optical frequency ratio measurements.
    }
    \label{fig:laser_setup_IS}
\end{figure}

The experimental setup is sketched in Fig.~\ref{fig:laser_setup_IS}. 
Two separate experiments are involved in the $\mathcal{R}$ measurements, one is based on a segmented linear radio-frequency (rf) Paul trap~\cite{pyka_high-precision_2014,keller_probing_2019}, which is used to trap even isotopes of Yb$^{+}$, and the other one is based on a ring style rf Paul trap~\cite{beaty1987simple} trapping $^{171}\mathrm{Yb}^{+}$. 
Two stable laser systems are employed for probing the $\alpha$ and $\gamma$ transitions. 
A laser at 822\,nm is frequency-doubled to interrogate the $\alpha$ transition near 411\,nm in the even isotopes. 
This laser is stabilized to an ultra-low expansion (ULE) optical cavity, obtaining short-term frequency stability of $5\times 10^{-16}$ at 10\,s averaging time~\cite{Keller2014simple}. 
The larger frequency shifts between the isotopes are bridged with an electro-optic modulator (EOM) in the infrared, while smaller frequency shifts are applied with a double-pass acousto-optic modulator (AOM) in the blue. 

A stable laser at 934\,nm is frequency-doubled to 467\,nm and continuously steered to be on resonance with the $\gamma$ transition of the $^{171}\mathrm{Yb}^{+}$ isotope~\cite{Lange2021improved}. Part of the infrared light is frequency offset via an EOM and used to stabilize the frequency of a second laser at 934\,nm. The frequency-doubled output of this laser is used for the interrogation of the $\gamma$ transitions in the even isotopes.
In both setups, the short-term frequency stability of the lasers is improved by locking to a cryogenic silicon cavity~\cite{Matei2017}. 
A frequency comb setup is used to perform the optical frequency ratio measurements. 

In the segmented, linear rf Paul trap setup used for the even isotopes, the quantization axis is defined by a static magnetic field of about $65$\,\textmu T oriented in the $xz$-plane at an angle of $\beta=26.8\,(4.0)$\textdegree\ with respect to the trap axis~$z$ (shown in Fig.~\ref{fig:experimental_setup}\,\textbf{a} in the main text), obtained from geometric measurements. The lasers to interrogate the $\alpha$ and the $\gamma$ transitions are aligned in the radial trapping direction $y$. Since the electronic states of the even Yb$^+$ isotopes do not contain sublevels with zero magnetic moment $m=0$, the $\alpha$ and $\gamma$ transitions are first-order magnetic-field sensitive. 
We employ active magnetic field stabilization~\cite{Fuerst2020coherent,Yeh2023robust} to increase the maximum coherent interrogation time for these transitions. 

AOMs are used to address pairs of Zeeman components with opposite magnetic field sensitivity and the center transition frequencies are obtained via averaging in post-processing. 
For the $\alpha$ and the $\gamma$ transitions, the $|{}^{2}\!S_{1\!/\!2}, m_{J}=\pm1/2\rangle\rightarrow|{}^{2}\!D_{5\!/\!2}, m_{J}=\pm5/2\rangle$ and the $|{}^{2}\!S_{1\!/\!2}, m_{J}=\pm1/2\rangle\rightarrow|{}^{2}\!F_{7\!/\!2}, m_{J}=\pm1/2\rangle$ transitions are interrogated, respectively.

The basic interrogation sequence starts with 5\,ms of laser cooling via the $^{2}\!S_{1\!/\!2}\!\rightarrow\!{}^{2}\!P_{1\!/\!2}$ electric dipole transition near 370\,nm, assisted by laser light near 935\,nm for repumping via the $^{2}\!D_{3\!/\!2}\!\rightarrow\!{}^{3}\![3/2]_{1\!/\!2}$ transition. It is followed by state preparation to either of the $|{}^{2}\!S_{1\!/\!2}, m_{J}=\pm1/2\rangle$ Zeeman sublevels via the $|{}^{2}\!S_{1\!/\!2}, m_{J}=\mp1/2\rangle\rightarrow|{}^{2}\!D_{5\!/\!2}, m_{J}=\pm3/2\rangle$ transitions, assisted by the $^{2}\!D_{5\!/\!2}\!\rightarrow\!{}^{2}\!P_{3\!/\!2}$ transition near 1650\,nm~\cite{roos2006designer}. A Rabi pulse of $t_{\pi}=2.1$\,ms ($t_{\pi}=\pi/\Omega_R$, where $\Omega_R$ is the Rabi frequency) or of about 50\,ms is employed to interrogate the $\alpha$ or the $\gamma$ transition. 
The maximum excitation probability is $P_{\alpha}=45$\% and $P_{\gamma}=80$\% for the $\alpha$ and the $\gamma$ transitions. 
After this, laser light near 370\,nm is applied for 2.5\,ms for state-selective fluorescence detection. Successful excitation results in the absence of fluorescence. After state detection, repumpers near wavelengths of 638\,nm and 1650\,nm for the $^{2}\!F_{7\!/\!2}\!\rightarrow\!{}^{1}[5/2]_{5/2}$ and the $^{2}\!D_{5\!/\!2}\!\rightarrow\!{}^{2}\!P_{3\!/\!2}$ transitions, respectively, are switched on to return the ion to the cooling cycle. A valid experimental cycle requires the detection of fluorescence after repumping. 

\begin{figure}[tb!]
    \centering
    \includegraphics[width=0.99\linewidth]{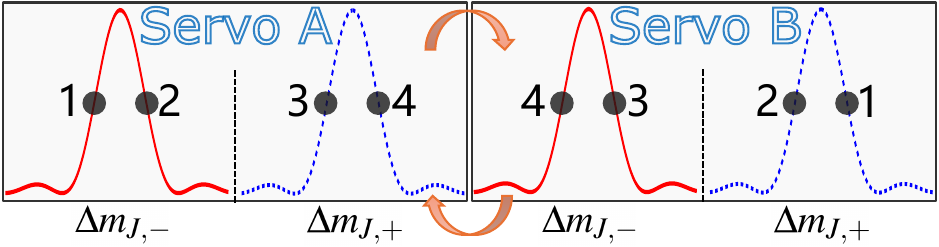}
    \caption{Schematic representation of the four-point servo lock used to determine the transition frequencies. The solid (red) and dashed (blue) curves correspond to the spectra of the $\Delta m_{J,\mp}:\, |^{2}\!S_{1\!/\!2}, m_{J}=\mp1/2\rangle\!\rightarrow\!|^{2}\!D_{5\!/\!2}, m_{J}=\mp5/2\rangle$ or the $\Delta m_{J,\mp}:\, |^{2}\!S_{1\!/\!2}, m_{J}=\pm1/2\rangle\!\rightarrow\!|^{2}\!F_{7\!/\!2}, m_{J}=\pm1/2\rangle$ transitions, respectively. The numbers denote the order of the interrogation frequencies. ``Servo A'' and ``Servo B'' are performed in an interleaved manner.}
    \label{fig:servo}
\end{figure}

This interrogation sequence is continuously repeated with 50 cycles at each point of a four-point servo lock. 
Here, the servo points run over fixed positive and negative detunings from each component of the Zeeman pair, as illustrated in Fig.~\ref{fig:servo}. 
After interrogating at all four points, the mean AOM frequency is updated to obtain the same excitation probability for equal positive and negative detunings. 
The $\gamma$ transition is additionally interrogated at a lower laser intensity using an independent servo. From the high- and low-intensity servos, the transition frequency that is free from probe light-induced ac-Stark shift is derived via linear extrapolation.

\section{Systematic and statistical uncertainties for the frequency measurements}\label{appen:sys_unc_freq_mea}

For each of the two transitions in the even isotopes, one systematic frequency shift dominates and causes most of the systematic uncertainty. 
The electric quadrupole shift stemming from the interaction of the quadrupole moment with the electric field gradient at the ion position dominates for the $\alpha$ transition. 
The electric quadrupole shift can be calculated from the electric field gradient, the electric quadrupole moment, and the angle $\beta$ between the symmetry axis of the gradient and the applied magnetic field. 
The electric field gradient is determined to be $dE_{z}/dz=3.1610\,(15)$\,V/mm$^{2}$ from an axial secular frequency of $\nu_{\text{ax}} = 211.97\,(5)\,$kHz. A potential electric field gradient of $0.2\,$V/mm$^{2}$ from stray electric fields is added in quadrature to the uncertainty of $dE_{z}/dz$. With the electric quadrupole moment of $12.5\,(4)\, a^{2}_{0}$~\cite{Tan2021precision}, we obtain an electric quadrupole shift of $-9.3\,(1.8)$\,Hz. For the $\gamma$ transition, this systematic shift is negligible due to a more than 400 times smaller quadrupole moment~\cite{Lange2020coherent}.

\begin{table*}[tb]
	\centering
	\caption[]{Leading systematic shifts $\delta \nu_{\alpha}$, $\delta \nu_{\gamma}$ and corresponding uncertainties $u_{B,\alpha}$, $u_{B,\gamma}$ of the $^{2}\!S_{1\!/\!2}\!\rightarrow\!{}^{2}\!D_{5\!/\!2}$ electric quadrupole ($\alpha$) and the $^{2}\!S_{1\!/\!2}\,\rightarrow\,^{2}\!F_{7\!/\!2}$ electric octupole ($\gamma$) transition for all even isotopes.}
  \begin{ruledtabular}
    \begin{tabular}{lrrrr}
    Systematic shift & $\delta \nu_{\alpha}$ [Hz] & $u_{B,\alpha}$ [Hz] & $\delta \nu_{\gamma}$ [Hz] & $u_{B,\gamma}$ [Hz]  \\ \hline \\ [-1.7ex]
        Electric quadrupole shift            & $-$9.3   & 1.8    & $-$0.016 & 0.003 \\ 
        Probe-light induced ac-Stark shift   & 0.002  & 0.001  & $-$9.2   & 11.1  \\
        Second-order Doppler shift           & $-$0.002 & 0.004  & $-$0.002 & 0.004 \\ 
        Second-order Stark shift             & $-$0.015 & 0.047  & -0.003 & 0.010 \\
        Blackbody radiation shift            & $-$0.26  & 0.13   & $-$0.045 & 0.001 \\  
        Servo error induced shift            & --     & 0.8    & --     & 0.1   \\ \hline \\ [-1.7ex]
        Total                 & $-$9.6 & 2.0  & $-$9.3 & 11.1  
    \end{tabular}%
  \end{ruledtabular}
	\label{tab:freq_systematic_uncertainties}
\end{table*}

The dominant systematic effect for the $\gamma$ transition is the probe-light induced ac-Stark shift. Even though Rabi excitation at two different laser intensities is employed to determine the transition frequency free from the ac-Stark shift, imperfect light path design caused the beam pointing to change with the rf power applied to the AOM. This shift effect is estimated by comparing the slope of the extrapolation to zero ac-Stark shift during the measurement campaign to that of a measurement where the change in the probe light power is realized independent of the AOM power. From the comparison, a maximum shift of $-9.2\,(11.1)$\,Hz is determined. In contrast, for the $\alpha$ transition, the probe-light induced ac-Stark shift even with a shorter pulse duration of $t_{\pi} = 2.1$\,ms is calculated from the differential polarizability in Ref.~\cite{Fuerst2020coherent} and causes a negligible shift of only $2\,(1)\,$mHz. 

In addition to the major systematic shifts discussed so far, Tab.~\ref{tab:freq_systematic_uncertainties} includes smaller shift effects and their uncertainties. These shift effects will be discussed in the following. 

The ion is confined with radial secular frequencies of about 600\,kHz and an axial secular frequency of around 200\,kHz and laser-cooled to the Lamb-Dicke regime reaching a temperature of 0.60\,(6)\,mK.
Minimization and measurements of the excess micromotion (EMM) using the photon correlation technique~\cite{keller_precise_2015} were performed only before and after the measurement campaign. 
Based on these measurements, we estimate the mean rf field causing EMM to be $\{E_{x}, E_{y}, E_{z}\} = \{100\,(500), 100\,(500), 150\,(50)\}$\,V/m. Here, the uncertainties are set to the maximum variations seen over the three months prior to the measurements. Both effects from uncompensated EMM and intrinsic micromotion due to the finite ion temperature are considered to calculate the second-order Doppler shift and the second-order Stark shift. 
The corresponding second-order Doppler shift is calculated as $-0.002\,(4)$\,Hz and the second-order Stark shift is $-0.015\,(47)$\,Hz and $-0.003\,(10)$\,Hz for the $\alpha$ and the $\gamma$ transition, respectively. The involved static differential polarizability for the $\alpha$ transition is found in Ref.~\cite{Fuerst2020coherent} and for the $\gamma$ transition in Ref.~\cite{Huntemann2016single}.

The blackbody radiation shift is calculated to be $-0.26\,(13)$\,Hz for the $\alpha$ transition and $-0.046\,(1)$\,Hz for the $\gamma$ transition considering thermal radiation with an effective temperature of $T_{C} = 297.1\,(1.4)\,$K perturbing the ion~\cite{Huntemann2016single,Fuerst2020coherent}. The temperature uncertainty assumed the maximum temperature difference measured at the chamber and the ion trap during the measurement campaign. 

Magnetic field variations can cause servo errors because of the long dead time and non-randomized four-point servo lock sequence~\cite{peik2005laser,Lindvall2023noise}. These are estimated to be 0.8\,Hz and 0.1\,Hz for the measurements of the $\alpha$ and the $\gamma$ transitions. By adding the uncertainties of the systematic shifts in quadrature, we determine the total systematic uncertainty for the $\alpha$ and the $\gamma$ transition frequencies to be $u_{B,\alpha}=2.0$\,Hz and $u_{B,\gamma}=11.1$\,Hz, respectively. 

The systematic uncertainty for the realization of the $\gamma$ transition in $^{171}\mathrm{Yb}^{+}$ has been evaluated to $1.7$\,mHz~\cite{sanner2019optical}. 
For the frequency ratio measurements, a gravitational redshift between the experimental setups for the even isotopes of Yb$^{+}$ and the $^{171}$Yb$^{+}$ of $-3.26\,(4)\times 10^{-17}$ has been taken into account. 

\begin{figure}[tbh!]
    \centering
    \includegraphics[width=0.99\linewidth]{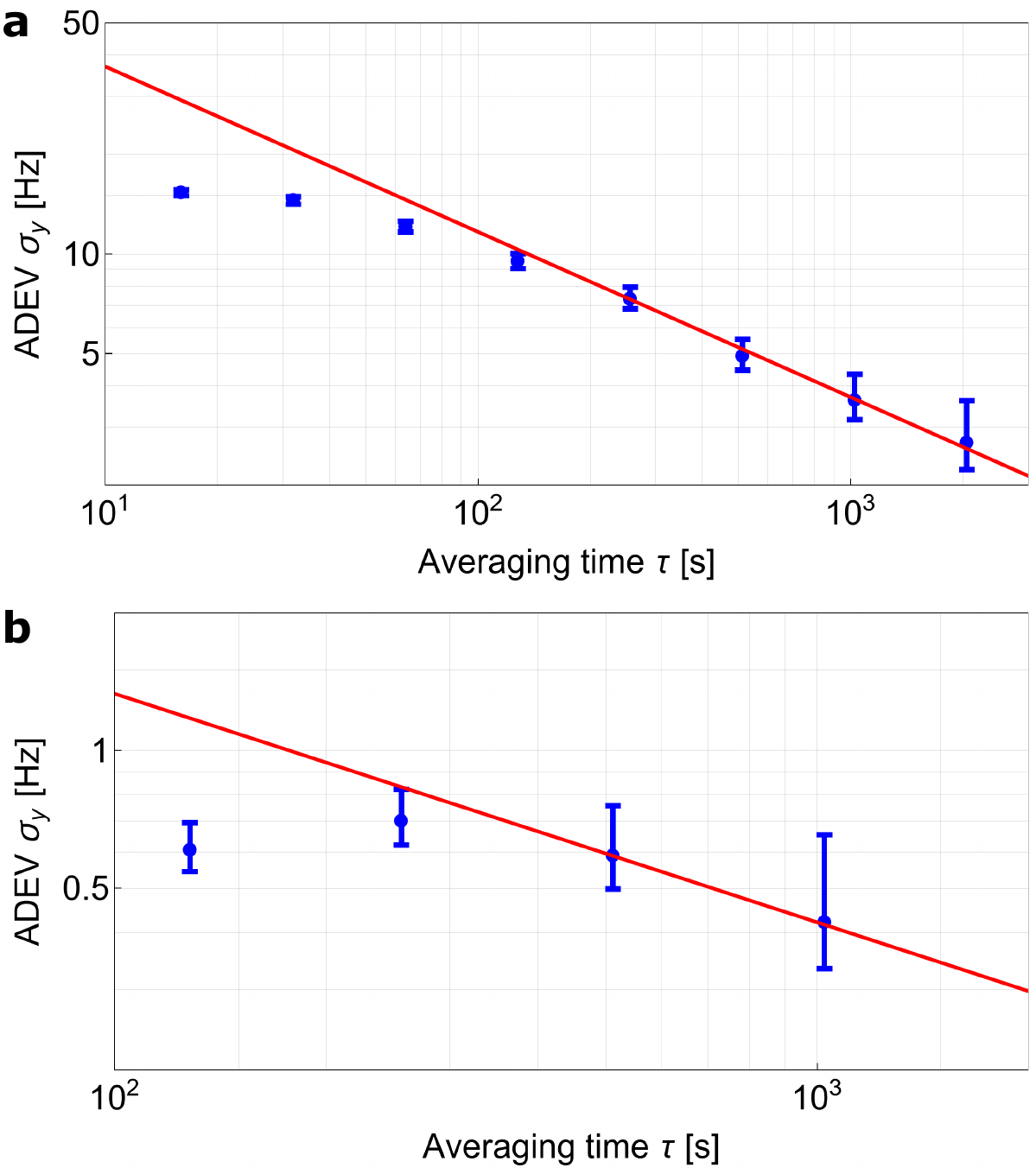}
    \caption{Allan deviations $\sigma_{y}$ of the $\alpha$ (\textbf{a}) and the $\gamma$ (\textbf{b}) transition frequency measurements in $^{168}$Yb$^{+}$. The fits to the data yield $\sigma_{\text{fit},\alpha} = 101.7\,(1.5)\,\mathrm{Hz}/\sqrt{\tau}$ and $\sigma_{\text{fit},\gamma} = 13.36\,(7)\,\mathrm{Hz}/\sqrt{\tau}$, agreeing with the instabilities expected from quantum projection noise.}
    \label{fig:ADEV_R}
\end{figure}

The statistical uncertainties $u_{A}$ from the measurements are listed in Tab.~\ref{tab:freq_statistical_uncertainties}. The values correspond to the last point in the Allan deviation $\sigma_{y}$ of the transition frequency measurements. Examples are shown in Fig.~\ref{fig:ADEV_R} for $\mathcal{R_{\alpha,\gamma}}$ of the $\alpha$ and the $\gamma$ transitions in $^{168}$Yb$^{+}$. 
The statistical uncertainty is smaller for the $\gamma$ transition because we operate at a different resolved linewidth. With pulsed Rabi spectroscopy, we resolve the $\alpha$ transition with good contrast at a linewidth of about 300 Hz, limited by the natural decay, while the resolved linewidth of the resonance of the $\gamma$ transition is below 20 Hz in these measurements.
The fit to the Allan deviation for the $\alpha$ transition frequency measurement yields $\sigma_{\text{fit},\alpha} = 101.7\,(1.5)\,\mathrm{Hz}/\sqrt{\tau}$,  and that of the Allan deviation for the $\gamma$ transition is $\sigma_{\text{fit},\gamma} = 13.36\,(7)\,\mathrm{Hz}/\sqrt{\tau}$. Both of the fits agree with the instability expected from quantum projection noise.

\begin{table}[tb]
	\centering
	\caption[]{Statistical uncertainties $u_{A,\alpha}$, $u_{A,\gamma}$ for determining the $\alpha$ and the $\gamma$ transition frequencies.}
 \begin{ruledtabular}
	\begin{tabular}{lrr}
            Isotope $A$ & $u_{A,\alpha}$ [Hz] & $u_{A,\gamma}$ [Hz]  \\ \hline \\ [-1.7ex]
		168  & 2.7 & 0.4 \\ 
		170  & 3.6 & 0.2 \\ 
		172  & 1.8 & 0.3 \\ 
		174  & 3.5 & 0.3 \\ 
		176  & 1.3 & 0.3 \\ 
	\end{tabular}
 \end{ruledtabular}
	\label{tab:freq_statistical_uncertainties}
\end{table}

The total uncertainties of the transition frequencies for each isotope are given by $\sigma \nu_{k} = \sqrt{u_{A,k}^{2}+u_{B,k}^{2}}$, with $k = \alpha, \gamma$. We determine the absolute frequencies of the $\alpha$ and the $\gamma$ transitions for all five stable even isotopes of Yb$^{+}$ with the frequency ratios, the literature value for the absolute frequency of the $\gamma$ transition in $^{171}$Yb$^{+}$~\cite{BIPM2022}, and the systematic shifts. Values are shown in Tab.~\ref{tab:abs_freq}.
For determining the isotope shifts, we do not consider any common mode rejection and take into account the full uncertainties as given in Tab.~\ref{tab:abs_freq}. The deviation $\delta \nu = \nu_{\textrm{TW}} - \nu_{\textrm{Ref.}}$ of the isotope shifts determined in this work (TW) and in prior publications are given in Tab.~1 in the main text for the $\alpha$ transition~\cite{Counts2020evidence}, and in Tab.~\ref{tab:diff_gammatransitions} for the $\gamma$ transition~\cite{Hur2022evidence}.

\begin{table}[tb]
	\centering
	\caption[]{Absolute frequencies $\nu_{\alpha,\gamma}$ of the $\alpha$ and the $\gamma$ transitions for all five stable even isotopes of Yb$^{+}$.}	
 \begin{ruledtabular}
	\begin{tabular}{lcc}
		Isotope $A$ & $\nu_{\alpha} - 729$\,THz [Hz] & $\nu_{\gamma} - 642$\,THz [Hz] \\ \hline \\ [-1.7ex]
		168  & 481\,090\,977\,315.9\,(3.4) & 108\,197\,800\,720.0\,(11.1) \\ 
		170  & 478\,911\,878\,447.9\,(4.1) & 112\,635\,960\,391.1\,(11.1) \\ 
		172  & 476\,867\,027\,166.9\,(2.7) & 116\,785\,150\,892.3\,(11.1) \\ 
		174  & 475\,283\,963\,017.6\,(4.0) & 119\,917\,471\,350.4\,(11.1) \\ 
		176  & 473\,774\,909\,821.9\,(2.4) & 122\,893\,863\,395.6\,(11.1) \\ 
	\end{tabular}
 \end{ruledtabular}
	\label{tab:abs_freq}
\end{table}

\begin{table}[tb]
	\centering
	\caption[]{Differences of the measured $\gamma$ transitions in this work compared to Ref.~\cite{Hur2022evidence} in units of Hz.}
	\begin{tabular}{l @{\qquad} c} 
        \toprule 
		 $A$ & $\Delta \nu_{\gamma}^{A,A+2}$ [Hz] \\
        \colrule
		168 & $-629\,(500)$~\cite{Hur2022evidence} \\ 
		170 & $-121\,(450)$~\cite{Hur2022evidence} \\ 
        172 & $-1142\,(500)$~\cite{Hur2022evidence} \\
		174 & $-445\,(480)$~\cite{Hur2022evidence} \\ 
        \botrule
	\end{tabular}
    \label{tab:diff_gammatransitions}
\end{table}

\section{King-plot analysis}

From the King plot shown in Fig.~\ref{fig:experimental_setup}\,\textbf{b,c} in the main text, we can extract the slope and offset, which correspond to the constants $F_{\gamma\alpha}$ and $K_{\gamma\alpha}$ in 
\begin{align}
    \mIS{\boldsymbol{\nu}}_{\gamma}\,\approx\,& 
    F_{\gamma\alpha}\mIS{\boldsymbol{\nu}}_{\alpha} + K_{\gamma\alpha}\boldsymbol{1}\,,
    \label{eq:A_kingrelation_vector}
\end{align}
which result in $-2.22131\,(5)$ and $0.033363\,(8)$\,Hz, respectively. We also resolve deviations from linearity on the average of 20.17\,(2)\,kHz. 
From the extracted slope, the nuclear mass-ratio uncertainties can be translated into frequency uncertainties. Here, the maximum translated frequency uncertainty is around 0.3\,Hz, stemming from the $(168,170)$ isotope pair, improving on the 28\,Hz uncertainty in Ref.~\cite{Hur2022evidence}.

\section{Nonlinearity decomposition}\label{app:NLD}

\begin{table}[tb]
    \centering
    \caption[]{Notation and references for the transitions used in the King-plot analysis and the nonlinearity decomposition plot. ``t.w.'' abbreviates ``this work''.}
    \begin{ruledtabular}
    \begin{tabular}{cccc}
        Notation    & Transition  & $\lambda$ [nm] & Refs.  \\ \hline \\ [-1.7ex]
        $\alpha_\mathrm{MIT}, \alpha_\mathrm{PTB}$    & $^{2}\!S_{1\!/\!2}\,\rightarrow\,^{2}\!D_{5\!/\!2}$ E2 in Yb$^{+}$  & 411  & \cite{Counts2020evidence,Hur2022evidence}, t.w. \\ 
        $\beta$     & $^{2}\!S_{1\!/\!2}\,\rightarrow\,^{2}\!D_{3\!/\!2}$ E2 in Yb$^{+}$  & 435  & \cite{Counts2020evidence}\\ 
        $\gamma_\mathrm{MIT}, \gamma_\mathrm{PTB}$    & $^{2}\!S_{1\!/\!2}\,\rightarrow\,^{2}\!F_{7\!/\!2}$ E3 in Yb$^{+}$  & 467   & \cite{Counts2020evidence,Hur2022evidence}, t.w.\\ 
        $\delta$    & ${}^{1}$$S_{0}\,\rightarrow\,{}^{3}$$P_{0}$ in Yb &  578   &\cite{Ono2022observation} \\ 
        $\epsilon$  & ${}^{1}$$S_{0}\,\rightarrow\,{}^{1}$$D_{2}$ in Yb &  361   & \cite{Figueroa2022precision}\\
    \end{tabular}
    \end{ruledtabular}
    \label{tab:transition_notation}
\end{table}

Given $n$ isotope pairs, $n-2$ King-plot nonlinearities can be resolved by projecting the isotope shift data on $n$ $n$-vectors. 
For simplicity, we use the same reference transition $\delta$ for all King plots entering our nonlinearity analysis and choose our basis in such a way that two basis vectors, $\mIS{\boldsymbol{\nu}}_{\delta}$ and $\boldsymbol{1}$ [see Eq.~\eqref{eq:A_kingrelation_vector} with $\alpha \rightarrow \delta$], lie within the plane of King linearity, whereas the other two, $
    \Lp \propto     (
        \mIS{\nu}_\delta^c - \mIS{\nu}_\delta^b, 
        \mIS{\nu}_\delta^a - \mIS{\nu}_\delta^d, 
        \mIS{\nu}_\delta^d - \mIS{\nu}_\delta^a, 
        \mIS{\nu}_\delta^b - \mIS{\nu}_\delta^c
    )$ and $
    \Lm \propto (
        \mIS{\nu}_\delta^d - \mIS{\nu}_\delta^b,
        \mIS{\nu}_\delta^a - \mIS{\nu}_\delta^c,
        \mIS{\nu}_\delta^b - \mIS{\nu}_\delta^d,
        \mIS{\nu}_\delta^c - \mIS{\nu}_\delta^a
    )$, 
normalized to 1\,Hz and with $a=(168,170)$, $b=(170,172)$, $c=(172,174)$, $d=(174,176)$,
are orthogonal to this plane and directly calculable from the experimental data. 
In this way, we can use the same basis for the nonlinearity decomposition of the experimental data and for the \emph{ab initio} predictions. The coordinates $(\lambda_+^{(\tau)}, \lambda_-^{(\tau)})$, $\tau\in\{\alpha_{\mathrm{PTB}}, \alpha_{\mathrm{MIT}}, \beta, \gamma_{\mathrm{PTB}}, \gamma_{\mathrm{MIT}}, \delta, \epsilon\}$, defined via 
\begin{equation}
    \mIS{\boldsymbol{\nu}}_\tau = 
        F_{\tau\delta} \mIS{\boldsymbol{\nu}}_\delta + K_{\tau\delta} \boldsymbol{1}  + \lambda_+^{(\tau)} \Lp + \lambda_-^{(\tau)} \Lm\,
    \label{eq:def_lambda_pm}
\end{equation}
characterize the nonlinearities in the isotope shift $\boldsymbol{\nu}_\tau$.

In the presence of one dominant source of King nonlinearity, the electronic coefficients $G^{(2)}, G^{(4)}, (\alpha_{\mathrm{NP}}/\alpha_{\mathrm{EM}}) D, \ldots$ [see Eq.~\eqref{eq:A_sotopeshift_all}] drop out of the ratio $\lambda_-^{(\tau)}/\lambda_+^{(\tau)}= \lambda_-/\lambda_+$, making $\lambda_-/\lambda_+$ a transition-independent quantity.
This motivates a linear fit through the origin of the $(\lambda_+,\lambda_-)$ plane to the six data points $(\lambda_+^{(\tau)}, \lambda_-^{(\tau)})$ derived for the six transitions listed in Tab.~\ref{tab:transition_notation}.
Indeed, the tension of the Yb isotope-shift data for King linearity can be drastically reduced by introducing a King nonlinearity satisfying $\lambda_-/\lambda_+=-1.777(14)$, leaving a residual nonlinearity at the level of $23\,\sigma$.

We compare this slope to predictions of individual nonlinearity terms in 
\begin{equation}
    \begin{aligned}
        \boldsymbol{\nu}_\alpha=& 
        F_\alpha \boldsymbol{\delta}\langle r^{2}\rangle 
        + K_\alpha \boldsymbol{w} 
        + G_{\alpha}^{(2)} \boldsymbol{\delta}\langle r^{2}\rangle^{2} 
        + G_{\alpha}^{(4)} \boldsymbol{\delta}\langle r^{4}\rangle \\
        &+ \frac{\alpha_{\mathrm{NP}}}{\alpha_{\mathrm{EM}}}D_{\alpha}\boldsymbol{h} + \ldots
    \end{aligned}
    \label{eq:A_sotopeshift_all}
\end{equation}
to identify the leading King-plot nonlinearity.
The new physics prediction for $\lambda_-/\lambda_+$ follows straightforwardly from Eqs.~\eqref{eq:A_sotopeshift_all} and \eqref{eq:def_lambda_pm}, and has a relative uncertainty of approximately $3 \times 10^{-7}$. 
The $\lambda_-/\lambda_+$ slope corresponding to the quadratic field shift $\drtwo^2$ [see Eq.~\eqref{eq:A_sotopeshift_all}, dotted line in Fig.~\ref{fig:NDP}\,\textbf{a} in the main text] can be derived from experimental data for $\drtwo{}$~\cite{ANGELI201369} with a relative uncertainty of approximately $2\%$.
If the new physics (quadratic field shift) term were to dominate the nonlinearity, the experimental data points would be expected to lie on the dash-dotted (dotted) line in Fig.~\ref{fig:NDP}\,\textbf{a} in the main text. We conclude that neither the new physics term nor the quadratic field shift term can be the dominant nonlinearity in the ytterbium King plot (Fig.~~\ref{fig:experimental_setup}\,\textbf{b,c} in the main text). 

Our \emph{ab initio} calculations of $\drfour$ fit better to data: In Fig.~\ref{fig:NDP}\,\textbf{a} in the main text we show the projections of our chiral effective field theory calculations using the 1.8/2.0~(EM) interactions with valence spaces VS1 (dashed, blue) and VS2 (dashed, orange). Both agree within uncertainties (the 68\% confidence interval is given by the gray band) with the linear fit to data, suggesting that $\drfour$ can explain the leading King nonlinearity in ytterbium. Our \emph{ab initio} calculations are discussed in more detail in Section~\ref{app:nuclear_structure}.

\section{\textit{Ab initio} nuclear structure calculations}\label{app:nuclear_structure}

The ground-state properties of atomic nuclei, such as \rtwo{} and \rfour{}, can be computed \textit{ab initio} using effective field theories for the strong interactions between nucleons and systematically improvable many-body methods to solve the many-body Schr\"odinger equation. Yb isotopes are challenging to describe \textit{ab initio} due to being heavy and open-shell, and our work leverages recent developments improving the treatment of three-nucleon forces in heavy nuclei~\cite{Miyagi2022no2b3n} and performing large-scale diagonalizations of the many-body Hamiltonian~\cite{Shimizu2021qvsm}.

Our calculations are performed with the valence-space IMSRG~\cite{Tsukiyama2011imsrg,Hergert2016imsrg,Stroberg2019vsimsrg}, which solves the many-body Schr\"odinger equation for a given input Hamiltonian $H$ via a unitary transformation to a block-diagonal form, $\overline{H} = U H U^\dagger$. The unitary transformation $U=e^\Omega$ is formulated with respect to a reference state $\ket{\Phi_0}$ for the system of interest and can as a result be efficiently approximated at the normal-ordered two-body level, the IMSRG(2). We solve the IMSRG(2) in a model space of 15 oscillator shells based on an underlying harmonic oscillator basis with frequency $\hbar\omega = 12\,\text{MeV}$, including three-nucleon forces with a truncation of $E_\text{3max} = 28$ made possible by Ref.~\cite{Miyagi2022no2b3n}. For the transformed Hamiltonian, an effective valence-space Hamiltonian is decoupled and subsequently diagonalized via large-scale shell-model methods.

We employ two nuclear Hamiltonians with two- and three-nucleon interactions derived within chiral effective field theory (EFT). The 1.8/2.0~(EM) Hamiltonian~\cite{Hebeler2011magic} is constructed from the N${}^3$LO nucleon-nucleon potential of Entem and Machleidt (EM)~\cite{Entem:2003ft}, transformed using the similarity renormalization group to a resolution scale of $\lambda = 1.8\,\fm^{-1}$, and N$^2$LO three-nucleon interactions with a regulator cutoff $\Lambda_\text{3N} = 2.0\,\fm^{-1}$. It is fit to few-body systems (up to \elem{He}{4}) and predicts ground-state energies, spectra, and differential radius trends well in medium-mass and heavy nuclei~\cite{Simonis:2017dny,Stroberg:2019bch,Hebeler:2020ocj,Arthuis:2024mnl}. The \dnnlogo{} Hamiltonian~\cite{Jiang2020deltan2logo} is constructed at N$^2$LO with explicit inclusion of $\Delta$ isobars in the EFT and is fit to few-body data and nuclear matter properties and optimized to reproduce bulk properties in medium-mass nuclei. Differences in results obtained by the two Hamiltonians reflect the underlying EFT uncertainty for nuclear forces.

We employ two valence spaces: VS1 with a \elem{Sn}{132} core and an active valence space consisting of \orbital{1}{g}{7}, \orbital{2}{d}{5}, \orbital{2}{d}{3}, \orbital{3}{s}{1}, \orbital{1}{h}{11} proton orbitals and \orbital{2}{f}{7}, \orbital{1}{h}{9}, \orbital{1}{i}{13}, \orbital{2}{f}{5}, \orbital{3}{p}{3}, \orbital{3}{p}{1} neutron orbitals; and VS2 with a \elem{Gd}{154} core and an active valence space consisting of \orbital{2}{d}{3}, \orbital{3}{s}{1}, \orbital{1}{h}{11} proton orbitals and \orbital{1}{h}{9}, \orbital{1}{i}{13}, \orbital{2}{f}{5}, \orbital{3}{p}{3}, \orbital{3}{p}{1} neutron orbitals. Varying the valence space allows us to assess some of the uncertainty due to the employed many-body approximation. Recent developments have made IMSRG calculations at the normal-ordered three-body level available~\cite{Heinz2021imsrg3}. These calculations can currently not be converged in Yb, but we performed restricted calculations to estimate the order of magnitude of truncated contributions in the IMSRG, another source of many-body uncertainties.

To compute \rtwo{} and \rfour{}, we evaluate the ground-state expectation values of the translationally invariant point-proton radius operators $R^2_p$ and $R^4_p$, with the definitions
\begin{align}
    R^2_p &=
    \sum_i \Big[
    (1 + \tau_i) \frac{1}{2Z}
    \Big( 1 - \frac{2}{A}\Big)
    +
    \frac{1}{A^2}
    \Big] \mathbf{r}_i^2
    \nonumber\\&\quad
    +\sum_{i<j} \Big[
    \frac{2}{A^2}
    - \frac{2}{AZ}
    \Big(1 + \frac{\tau_i + \tau_j}{2} \Big)
    \Big] \mathbf{r}_i \cdot \mathbf{r}_j\,,
    \\ R^4_p &=
    \sum_i \Big[
    (1 + \tau_i) \frac{1}{2Z}
    \Big( 1 - \frac{4}{A}\Big)
    \Big] \mathbf{r}_i^4
    \nonumber\\&\quad
    +\sum_{i<j}
    \frac{-2}{AZ}
    \big[
    (1 + \tau_i) \mathbf{r}_i^2
    + (1 + \tau_j) \mathbf{r}_j^2
    \big]
    \mathbf{r}_i \cdot \mathbf{r}_j + \mathcal{O}(A^{-2})\,.
\end{align}
We neglect the indicated higher-order contributions in $R^4_p$, which would also include three- and four-body parts but are strongly suppressed. For \rtwo{}, we also include the spin-orbit~\cite{Ong2010spinorbit} and relativistic Darwin-Foldy~\cite{Friar1997darwinfoldy} corrections and account for the finite size of nucleons.

\begin{figure}[th!]
   \centering
  \includegraphics[width=1.0\linewidth]{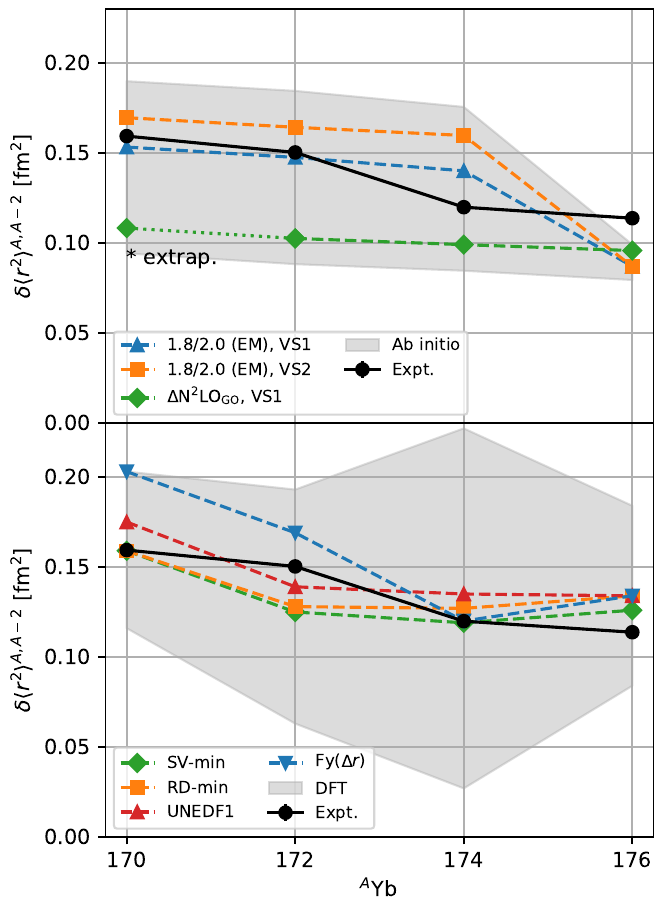} 
	\caption{Experimental $\drtwo^{A,A-2}$ values~\cite{ANGELI201369} are compared to nuclear theory predictions from our \textit{ab initio} calculations [top, 1.8/2.0~(EM), VS1 and VS2; $\Delta\text{N}^2\text{LO}_\text{GO}$, VS1] and from density functional theory calculations~\cite{Hur2022evidence} [bottom, SV-min, RD-min, UNEDF1, $\text{Fy}(\Delta r)$] for $A\in \{170,172,174,176\}$. The gray bands give estimated uncertainties of the theory results. 
        }
	\label{fig:rp2_diffs}
\end{figure}

Our computed values of \rtwo{}, \rfour{}, \drtwo{}, and \drfour{} are given in Tabs.~\ref{tab:nuclth_abs} and \ref{tab:nuclth_diff}. We note that we obtained nonphysical results for \elem{Yb}{168} with the \dnnlogo{} Hamiltonian, with an inversion of proton \orbital{2}{d}{3} and \orbital{1}{h}{11} single-particle orbitals at the Hartree-Fock level relative to \elem{Yb}{170-176}. For the \drtwo{} and \drfour{} values in Tab.~\ref{tab:nuclth_diff}, we obtain extrapolated values for \dnnlogo{} based on the differences observed for (172,170) and (170,168) for the 1.8/2.0~(EM) Hamiltonian in the same valence space. Our results are compared with experimental measurements of \drtwo{} from Ref.~\cite{ANGELI201369} and our extraction of \drfour{} trends in Fig.~\ref{fig:rp2_diffs} and Fig.~\ref{fig:rp4comparison_dft}, respectively.

To assess theoretical uncertainties, we employ a correlated statistical uncertainty model. For the propagation of uncertainties in \drtwo{} and \drfour{} to King-plot analyses, it is essential to account for correlations across isotope pairs. Errors due to truncations in \textit{ab initio} nuclear theory are very systematic in nature, producing similar errors in neighboring isotopes. As a result, strong correlations arise, and we attempt to quantify and account for these in the following.

We assume that a prediction for given observable $\mathbf{o} = (o^{(170,168)}, o^{(172,170)}, o^{(174,172)}, o^{(176,174)})$ (true value unknown) can be understood as
\begin{equation}
    \mathbf{o} = \mathbf{\tilde{o}} + \boldsymbol{\varepsilon}_\text{EFT} + \boldsymbol{\varepsilon}_\text{VS} + \boldsymbol{\varepsilon}_\text{MB}\,,
\end{equation}
where $\mathbf{\tilde{o}}$ is the approximate prediction
and $\boldsymbol{\varepsilon}_\text{EFT}$, $\boldsymbol{\varepsilon}_\text{VS}$, and $\boldsymbol{\varepsilon}_\text{MB}$ are errors made due the necessary truncations in the EFT for the Hamiltonians, the valence space, and the many-body method, respectively.
We choose
\begin{equation}
    \mathbf{\tilde{o}} = 0.75\, \mathbf{o}_{\nuclearmodelone} + 0.25\, \mathbf{o}_{\nuclearmodelthree}
\end{equation}
based on the reproduction of experimental \drtwo{} values. Additionally, we model the EFT, valence-space, and many-body errors as random variables distributed as multivariate normal distributions,
\begin{align}
    \boldsymbol{\varepsilon}_\text{EFT} &\sim \mathcal{N}(0, \Sigma_\text{EFT})\,, \\
    \boldsymbol{\varepsilon}_\text{VS} &\sim \mathcal{N}(0, \Sigma_\text{VS})\,, \\
    \boldsymbol{\varepsilon}_\text{MB} &\sim \mathcal{N}(0, \Sigma_\text{MB})\,.
\end{align}
We use full covariance matrices $\Sigma$ to allow us to consider correlations across isotope pairs in our error model, which we discuss below.

We estimate the variances [i.e., $\boldsymbol{\sigma}_i^2 = \text{diag}(\Sigma_i)$] for our EFT and valence-space errors as
\begin{align}
    \boldsymbol{\sigma}_\text{EFT}^2 &= (\mathbf{o}_{\nuclearmodelone} - \mathbf{o}_{\nuclearmodelthree})^2\,, \\
    \boldsymbol{\sigma}_\text{VS}^2 &= (\mathbf{o}_{\nuclearmodelone} - \mathbf{o}_{\nuclearmodeltwo})^2\,.
\end{align}

To estimate $\boldsymbol{\sigma}_\text{MB}^2$, we performed IMSRG(3)-$N^7$ calculations for \elem{Yb}{172}~\cite{Heinz2021imsrg3}. IMSRG(3) calculations cannot currently be converged in Yb, but by including restricted three-body operators in the calculations, we gain insight into the magnitude of the IMSRG(3) corrections. Moreover, calculations in carbon and calcium have shown that the three-body corrections are strongly correlated across systems (with a correlation coefficient $r\geq0.99$). We find the IMSRG(3) corrections for \rtwo{} and \rfour{} in \elem{Yb}{172} to be on the order of $0.06\:\fm^2$ and $2\:\fm^4$, respectively. Assuming a size extensive scaling of radius corrections and exploiting the strong correlation between neighboring isotopes, we find small many-body uncertainties $\boldsymbol{\sigma}_\text{MB}^2$ for the differential quantities \drtwo{} and \drfour{}. All employed standard deviations are given in Tab.~\ref{tab:nuclth_err}.

Finally, the covariance between $\varepsilon_i^{(A,A-2)}$ and $\varepsilon_i^{(B,B-2)}$ ($i \in \{\text{EFT}, \text{VS}\}$) is estimated as
\begin{equation}
    \text{cov}(\varepsilon_i^{(A,A-2)}, \varepsilon_i^{(B,B-2)}) = r_i^{(A - B)/2} \sigma_i^{(A,A-2)}
    \sigma_i^{(B,B-2)}\,,
\end{equation}
assuming an exponentially decaying correlation with a correlation $r_i$. Based on tests of different correlation values against our computed values, we found $r_\text{EFT}=0.99$ and $r_\text{VS}=0.97$. For the many-body uncertainties, we conservatively assumed no correlations.

We emphasize that these uncertainties and the underlying variances are not meant to be interpreted statistically but rather as an expert assessment of underlying parameters in our model to estimate correlated uncertainties and propagate those to King-plot nonlinearity analyses. We find this model to work well for \drtwo{} when compared to experimental values and expect it to work similarly well for \drfour{}.

\begin{table}[th!]
	\centering
	\caption[]{Absolute values of \rtwo{} and \rfour{} for the 1.8/2.0~(EM) Hamiltonian using valence spaces VS1 and VS2 and for the \dnnlogo{} Hamiltonian using VS1. Missing results for \elem{Yb}{168} with the \dnnlogo{} Hamiltonian are discussed in the main text.}
 \begin{ruledtabular}
	\begin{tabular}{c|ccc}
		  \multirow{3}{*}{\makecell{Isotope \\ $A$}} & \multicolumn{3}{c}{\rtwo{} [$\fm^2$]} \\
            &  \multicolumn{2}{c}{1.8/2.0~(EM)} & \dnnlogo{} \\
            & \hphantom{**}VS1\hphantom{**} & \hphantom{**}VS2\hphantom{**} & \hphantom{**}VS1\hphantom{**} \\
            \\[-2.3ex]\hline\\[-2.3ex]
		168  &  24.233 &  24.287 & (*) \\
            170  &  24.387 & 24.457 & 26.609 \\
            172  &  24.534 &  24.621 & 26.711 \\
            174  &  24.674 &  24.781 & 26.810 \\
            176  &  24.761 & 24.868 & 26.906  \\
            \\[-2.3ex]\hline\\[-2.3ex]
		  \multirow{3}{*}{\makecell{Isotope \\ $A$}} & \multicolumn{3}{c}{\rfour{} [$\fm^4$]}  \\
            & \multicolumn{2}{c}{1.8/2.0~(EM)} & \dnnlogo{} \\
            & VS1 & VS2 & VS1  \\
		\\[-2.3ex]\hline\\[-2.3ex]
		168   & 710.4 & 719.5 & (*) \\
            170   & 719.3 & 728.9 & 850.9 \\
            172   & 727.8 & 738.4 & 857.4 \\
            174   & 736.2 & 747.8 & 863.6 \\
            176   & 743.3 & 755.5 & 869.6 \\
	\end{tabular}
\end{ruledtabular}
	\label{tab:nuclth_abs}
\end{table}

\begin{table}[th!]
	\centering
	\caption[]{\drtwo{} and \drfour{} values for the 1.8/2.0~(EM) Hamiltonian using valence spaces VS1 and VS2 and for the \dnnlogo{} Hamiltonian using VS1. Extrapolated results~(*) for $(A,A')=(170,168)$ for the \dnnlogo{} Hamiltonian are discussed in the main text.}
  \begin{ruledtabular}
	\begin{tabular}{c|ccc}
		  \multirow{3}{*}{\makecell{Isotope pair \\ $A,A'$}} & \multicolumn{3}{c}{$\drtwo{}^{(A,A')}$ [$\fm^2$]} \\
            &  \multicolumn{2}{c}{1.8/2.0~(EM)} & \dnnlogo{} \\
            & \hphantom{**}VS1\hphantom{**} & \hphantom{**}VS2\hphantom{**} & \hphantom{**}VS1\hphantom{**} \\
		\\[-2.3ex]\hline\\[-2.3ex]
            (170,168)  &  0.153 & 0.170 & \hphantom{(*)}0.108(*) \\
            (172,170)  &  0.148 &  0.164 & 0.103 \\
            (174,172)  &  0.140 &  0.160 & 0.099 \\
            (176,174)  &  0.087 & 0.087 & 0.096  \\
            \\[-2.3ex]\hline\\[-2.3ex]
		  \multirow{3}{*}{\makecell{Isotope pair \\ $A,A'$}} & \multicolumn{3}{c}{$\drfour{}^{(A,A')}$ [$\fm^4$]}  \\
            & \multicolumn{2}{c}{1.8/2.0~(EM)} & \dnnlogo{} \\
            & VS1 & VS2 & VS1  \\
		\\[-2.3ex]\hline\\[-2.3ex]
            (170,168)   & 8.85 & 9.48 & \hphantom{(*)}6.81(*) \\
            (172,170)   & 8.51 & 9.48 & 6.47 \\
            (174,172)   & 8.35 & 9.34 & 6.21 \\
            (176,174)   & 7.11 & 7.70 & 6.00 \\
	\end{tabular}
 \end{ruledtabular}
	\label{tab:nuclth_diff}
\end{table}

\begin{table}[th!]
	\centering
	\caption[]{Assessed standard deviations for \drtwo{} and \drfour{} EFT, valence-space, and many-body errors.}	
 \begin{ruledtabular}
	\begin{tabular}{c|ccc}
		  \multirow{3}{*}{\makecell{Isotope pair \\ $A,A'$}} & \multicolumn{3}{c}{$\drtwo{}^{(A,A')}$ [$\fm^2$]} \\
            & \multirow{2}{*}{\makecell{$\sigma_\text{EFT}^{(A,A')}$}} & \multirow{2}{*}{\makecell{$\sigma_\text{VS}^{(A,A')}$}} & \multirow{2}{*}{\makecell{$\sigma_\text{MB}^{(A,A')}$}} \\
            & & & \\
		\\[-2.3ex]\hline\\[-2.3ex]
            (170,168)  &  0.045 & 0.016 & 0.005 \\
            (172,170)  &  0.045 &  0.017 & 0.005 \\
            (174,172)  &  0.041 &  0.020 & 0.005 \\
            (176,174)  &  0.009 & 0.001 & 0.005  \\
            \\[-2.3ex]\hline\\[-2.3ex]
		  \multirow{3}{*}{\makecell{Isotope pair \\ $A,A'$}} & \multicolumn{3}{c}{$\drfour{}^{(A,A')}$ [$\fm^4$]}  \\
            & \multirow{2}{*}{\makecell{$\sigma_\text{EFT}^{(A,A')}$}} & \multirow{2}{*}{\makecell{$\sigma_\text{VS}^{(A,A')}$}} & \multirow{2}{*}{\makecell{$\sigma_\text{MB}^{(A,A')}$}} \\
            & & & \\
		\\[-2.3ex]\hline\\[-2.3ex]
            (170,168)   & 2.04 & 0.63 & 0.15 \\
            (172,170)   & 2.04 & 0.97 & 0.16 \\
            (174,172)   & 2.14 & 0.98 & 0.16 \\
            (176,174)   & 1.11 & 0.58 & 0.16 \\
	\end{tabular}
 \end{ruledtabular}
	\label{tab:nuclth_err}
\end{table}

\section{\emph{Ab initio} atomic calculations of electronic coefficients}
\label{app:ambit}

The dependence of isotope shifts on the nuclear structure parameters considered in this work is quantified by the electronic coefficients $F$, $K$, $G^{(2)}$, $G^{(4)}$, and $D$ of Eq.~\eqref{eq:A_sotopeshift_all}. 
These can be calculated from atomic theory, however the accuracy is severely limited because of the strong many-body correlations. Worse, because parameters that depend on nuclear size and shape all depend on the wavefunction at the nucleus, the coefficients $F$ and $G^{(4)}$ (and $D$ at large $m_\phi$) are nearly proportional to each other for different transitions. That means that quantities such as $G^{(4)}_{\gamma\alpha} = G^{(4)}_{\gamma} - F_{\gamma\alpha} G^{(4)}_{\alpha}$ that appear in the King linearity [see Eq.~\eqref{eq:A_kingrelation_vector}] contain strong cancellations and therefore have large errors. A key advantage of the approach outlined in Section~\ref{app:atomic_dr4} is that the changes in $\drfour$ are extracted with the accuracy of $G^{(4)}_\alpha$ rather than $G^{(4)}_{\gamma\alpha}$.

We have used AMBiT~\cite{kahl19cpc} to perform a particle-hole configuration interaction (CI) calculation of these parameters for both Yb and Yb$^+$. For both atom and ion, we start with a self-consistent Dirac-Fock calculation for the closed-shell core with 68 electrons (up to the filled $4f^{14}$ shell). We then generate $6s$, $6p$, and $5d$ valence orbitals by solving the Dirac-Fock equations in the potential of this core. An orbital basis is then constructed by multiplying these orbitals with simple polynomials and orthogonality~\cite{bogdanovich83spc,kozlov96jpb}. The nuclear charge is represented by a Fermi-Dirac distribution.

For the case of transitions in Yb$^+$ ($\alpha$, $\beta$, $\gamma$), we freeze the core $5sp4d$ orbitals\,\footnote{In this notation the integers refer to the highest principal quantum numbers of the orbitals that follow. Here it refers to $1s-5s$, $2p-5p$, and $3d-4d$.}. The $4f$ shell is treated as a (hole) valence shell. We allow single and double electron excitations up to $8spdf$ from leading configurations $6s$, $5d$, $6p$, $4f^{-1}\;6s^2$, $4f^{-1}\;6s\;5d$, $4f^{-1}\;5d^2$, and $4f^{-1}\;6p^2$, with the additional restriction that a maximum of two holes in the $4f$ shell were allowed. Configuration state functions, with defined angular momentum $J$ and projection $M=J$ were formed from these configurations, and all were included in the CI Hamiltonian, which was then diagonalized to obtain level energies and wavefunctions. Coefficients calculated with this method are listed as ``Fiducial'' in Tab.~\ref{tab:electronic-coefficients}.

\begin{table*}[th!]
\begin{ruledtabular}
\begin{tabular}{c|cccc|cccc|cc}
\multirow{3}{*}{\makecell{Transition \\ $\tau$}}&
\multicolumn{4}{c|}{$F_\tau$ [GHz/fm$^2$]} 
&
\multicolumn{4}{c|}{$G_\tau^{(2)}$ [MHz/fm$^4$]}
&
\multicolumn{2}{c}{$G_\tau^{(4)}$ [MHz/fm$^4$]}\\
&
\multicolumn{2}{c}{Fiducial}
&
\multicolumn{2}{c|}{Core holes}
&
\multicolumn{2}{c}{Fiducial}
&
\multicolumn{2}{c|}{Core holes} 
&
\multicolumn{1}{c}{Fiducial}
&
\multicolumn{1}{c}{Core holes}\\
 &
\multicolumn{1}{c}{FF} & \multicolumn{1}{c}{$\langle F\rangle$} & 
\multicolumn{1}{c}{FF} & \multicolumn{1}{c|}{$\langle F\rangle$} & 
\multicolumn{1}{c}{FF} & \multicolumn{1}{c}{$\langle G^{(2)} \rangle$} & 
\multicolumn{1}{c}{FF} & \multicolumn{1}{c|}{$\langle G^{(2)} \rangle$} & 
\multicolumn{1}{c}{$\langle G^{(4)}\rangle$} 
& \multicolumn{1}{c}{$\langle G^{(4)}\rangle$} 
\\
\hline
$\alpha$  & -14.69  &  -13.09  &  -16.22  &  -16.14  &   81.8  &   72.4  &   90.3  &   89.2  &   8.76  &  10.80   \\
$\beta$   & -14.91  &  -13.07  &  -16.51  &  -16.45  &   83.0   &   72.3  &   91.8  &   90.9  &   8.75  &  11.00  \\
$\gamma$  &  37.78  &   29.09  &   33.24  &   45.45  & -209.7  & -160.8  & -184.5  & -250.7  & -19.47  & -30.38 \\
$\delta$  &  -9.73  &  -10.62  &   -9.82  &  -11.3   &   54.2  &   58.8  &   54.9  &   62.2  &   7.12  &   7.57  \\
$\epsilon$ & -13.54  &  -14.82  &  -13.69  &  -15.55  &   75.3  &   81.9  &   76.3  &   85.9  &   9.91  &  10.41 
\end{tabular}
\end{ruledtabular}
\caption{Electronic field shift coefficients $F_\tau$, quadratic field shift coefficients $G_\tau^{(2)}$, and quartic shift coefficients $G_\tau^{(4)}$ of the transitions $\tau$ defined in Tab.~\ref{tab:transition_notation}. }
\label{tab:electronic-coefficients}
\end{table*}

For the fiducial calculations of Yb transitions ($\delta$ and $\epsilon$) we can keep the $4f$ shell within the frozen core and treat the atom as having two valence electrons above closed shells. We use the same Dirac-Fock potential as in Yb$^+$ (in the case of Yb this is a $V^{N-2}$ potential). We expand the virtual basis up to $12spdf$ and include all possible configurations of the two valence electrons within this basis.

To calculate electronic coefficients, we used both a finite field approach and an operator approach. In the finite field approach the nuclear radius is varied around the physical point and the coefficient is extracted by numerical derivative. For example, $F_\tau = d\omega_\tau/d\langle r^2\rangle$ where $\omega_\tau$ is the calculated transition frequency and $r$ is the nuclear root-mean-square radius. This method is listed as ``FF'' in Tab.~\ref{tab:electronic-coefficients}. The operator approach simply evaluates the expectation value of an effective operator at the physical point, for example, $F_\tau = \langle \delta V_\textrm{nuc} \rangle_i$, where $\delta V_\textrm{nuc}$ is the change in the nuclear potential caused by $\drtwo$. This method is listed as ``$\langle O \rangle$'' in Tab.~\ref{tab:electronic-coefficients}. We see a significant difference between the two, which is expected since the operator approach does not include core relaxation. Nevertheless, we include it here as a ``worst case'' indication of theoretical uncertainty.

To provide a more reliable uncertainty measure, we have performed very large CI calculations by including, in addition, all configurations that can be formed by taking single electron excitations from core orbitals and up to two excitations of valence electrons from the leading configurations. Here we are inspired by~\cite{kozlov22pra} who suggest using single core excitations to treat valence and core-valence correlations more efficiently. We also expect that by unfreezing the core, the calculations of short-range matrix elements could be improved. The results of our calculations that include these additional core-excited configurations are listed in Tab.~\ref{tab:electronic-coefficients} as ``Core holes''. We see that generally, the difference between finite field and operator approaches to the matrix elements is smaller when using core holes, reflecting that the relaxation of the core is partially included using this method. The exception is the $\gamma$ transition for which we see a large difference: this transition directly involves a change in the occupancy of the $4f$ valence state, and the relaxation of this shell is large. Again we stress that our core hole approach is used as a measure of the theoretical uncertainty associated with having frozen cores in our fiducial CI method.

\begin{figure}
    \centering
    \includegraphics[width=\linewidth]{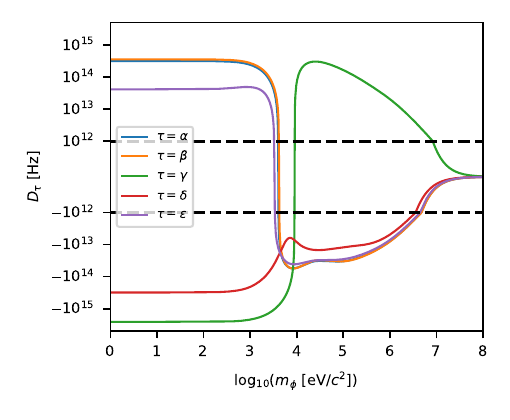}
    \caption{Electronic coefficients $D_\tau$, $\tau \in \{\alpha, \beta, \gamma, \delta,\epsilon\}$, introduced in Eq.~(1) of the main text, as a function of the mass $m_\phi$ of the new boson. The differences between $D_\alpha$ and $D_\beta$ are barely visible. The linear thresholds of the symmetric logarithmic plot are indicated with black dashed lines. Note the difference in convention with respect to Fig.~S14 in Ref.~\cite{Hur2022evidence}.}
    \label{fig:D_mphi}
\end{figure}

The $D$ coefficients that enter the bounds on new physics via the generalized King plot~\cite{Berengut:2020itu} are computed as described above (see also Ref.~\cite{Hur2022evidence}), with the addition of a finite Yukawa potential $\lambda e^{-m_\phi r c/\hbar}/r$. For each new boson mass $m_\phi$ we repeat the calculation with several values of $\lambda$ and take the numerical derivative.
Figure~\ref{fig:D_mphi} shows the values of the $D$ coefficients $D_\tau$, $\tau \in \{\alpha, \beta,\gamma,\delta,\epsilon\}$, as a function of the mass $m_\phi$ of the new boson, whereas Table~\ref{tab:mphi-Dcoeffs} provides a list of $D$ coefficients for all five transitions at selected values of $m_\phi$. Note the difference in convention with respect to Ref.~\cite{Hur2022evidence}, which does not normalize the new physics coupling $\alpha_{\mathrm{NP}}$ by the fine structure constant $\alpha_{\mathrm{EM}}$. The largest uncertainty in our calculation is due to the neglect of core polarization. Generally this will increase the size of the $D$ coefficients by up to 20\%, which is an estimate of our error. We found that not including core-polarization improved numerical stability and gave more conservative bounds on new physics.

\begin{table}[]
\begin{ruledtabular}
\begin{tabular}{l|c|r} 
    \phantom{*}$m_\phi$ [$\text{eV}/c^2$]\phantom{*} & \phantom{**}$i$\phantom{**} & \phantom{*} $D_i$ [Hz] \phantom{00000} \\
    \hline
        $\phantom{00}1$  
                & $\alpha$ & 314941096838754.4\phantom{000000}\\
                & $\beta$ & 354896899928149.1\phantom{000000} \\
                & $\gamma$ & -2571409426668460.5\phantom{000000}\\
                & $\delta$ &  -312730940277077.6\phantom{000000} \\
                & $\epsilon$ &  41363423961625.695\phantom{0000} \\  
        \hline
        $\phantom{0}10^2$  
                & $\alpha$ & 313713223754497.7\phantom{000000}\\
                & $\beta$ & 353538697268367.2\phantom{000000} \\
                & $\gamma$ & -2563969680971037.5\phantom{000000}\\
                & $\delta$ &  -310519664492137.25\phantom{00000}\\
                & $\epsilon$ &  41801954702786.555\phantom{0000}\\
        \hline
        $\phantom{0}10^4$  
                & $\alpha$ & -53135793622491.055\phantom{0000}\\
                & $\beta$ & -53749008165471.24\phantom{00000} \\
                & $\gamma$ &68784591261563.305\phantom{0000}\\
                & $\delta$ &  -8440014196795.924\phantom{0000}\\
                & $\epsilon$ & -40882557593630.83\phantom{00000} \\
        \hline
        $\phantom{0}10^6$  
                & $\alpha$ &   -7939928663244.745\phantom{0000}\\
                & $\beta$ &   -8128695194140.711\phantom{0000}\\
                & $\gamma$ & 20539381981880.473\phantom{0000}\\
                & $\delta$ &   -4930494151686.425\phantom{0000}\\
                & $\epsilon$ &   -7245790266413.627\phantom{0000} \\
        \hline
        $\phantom{0}10^8$  
                & $\alpha$ &   -4923184224.180642\phantom{0} \\
                & $\beta$ &   -5007700585.950998\phantom{0} \\
                & $\gamma$ & 12117018040.460941\phantom{0} \\
                & $\delta$ &   -3251693925.4536247 \\
                & $\epsilon$ &   -4526134700.21715\phantom{00} \\
    \end{tabular}
    \end{ruledtabular}
    \caption{Electronic $D$ coefficients of transitions $i$ (see Eq.~(1) in the main text) for selected values of the mediator mass $m_\phi$, calculated using AMBiT~\cite{kahl19cpc}.}
    \label{tab:mphi-Dcoeffs}
\end{table}

\section{Determining the evolution of $\delta \langle r^4\rangle$ along the isotope chain from King-plot data}\label{app:atomic_dr4}

Assuming that the leading source of nonlinearity in the King plot is \drfour{}, we can use isotope-shift measurements to extract information on this quantity. 
We proceed in two steps: 
Using the precise nuclear mass and frequency measurements presented in this work, we construct the quantities
\begin{align}
    \hat{\nu}_\tau^{a,r} = &\nu_\tau^{a} - \frac{w^{a}}{w^{r}} \nu_\tau^{r}\,,
    \label{eq:dotnu}
\end{align}
where $\tau$ is the transition index, $r=(174,176)$ denotes the reference isotope pair, and $a \in \{(168, 170), (170, 172), (172, 174)\}$ is an isotope pair index. 
Approximating Eq.~\eqref{eq:A_sotopeshift_all} (with $\alpha\to \tau$) by
\begin{align}
    \nu_\tau^{a}\, \approx &\,F_\tau \drtwo^{a} + K_\tau w^{a} + G_\tau^{(4)}\drfour^{a}\,,
    \label{eq:isotopeshift_w_dr4}
\end{align}
we obtain
\begin{align}
    \hat{\nu}_\tau^{a,r}\approx F_\tau D^{a,r} + G_\tau^{(4)} Q^{a,r}\,,
\end{align}
where the isotope-pair dependent quantity 
\begin{align}
    D^{a,r} = & \drtwo^{a} - \frac{w^{a}}{w^{r}} \drtwo^{r}\,,
\end{align}
can be constructed from the charge radius measurements tabulated in Ref.~\cite{ANGELI201369}, and
\begin{align}
    Q^{a,r}= & \drfour^{a} - \frac{w^{a}}{w^{r}} \drfour^{r}
    \label{eq:Qdef}
\end{align}
describes the evolution of the nuclear deformation parameter $\drfour$ along the isotope chain.
Following the King-plot approach, we perform a linear fit in isotope pair space, i.e., we fit $F_\tau$ to
\begin{align}
    \hat{\boldsymbol{\nu}}_\tau^r\approx F_\tau \mathbf{D}^r\,,
    \label{eq:dr4_fi_fit}
\end{align}
where $ \mathbf{x}^r = (x^{a,r},x^{b,r},x^{c,r},x^{d,r})$, $x\in\{\hat{\nu}_\tau, D\}$.
In this way, the precision of the isotope-shift measurements and nuclear mass measurements is exploited and field shift coefficients $F_\tau$ with relative uncertainties at the level of $\approx 0.3\%$ can be obtained.

\begin{figure}[th!]
   \centering
 \includegraphics[width=1.0\linewidth]{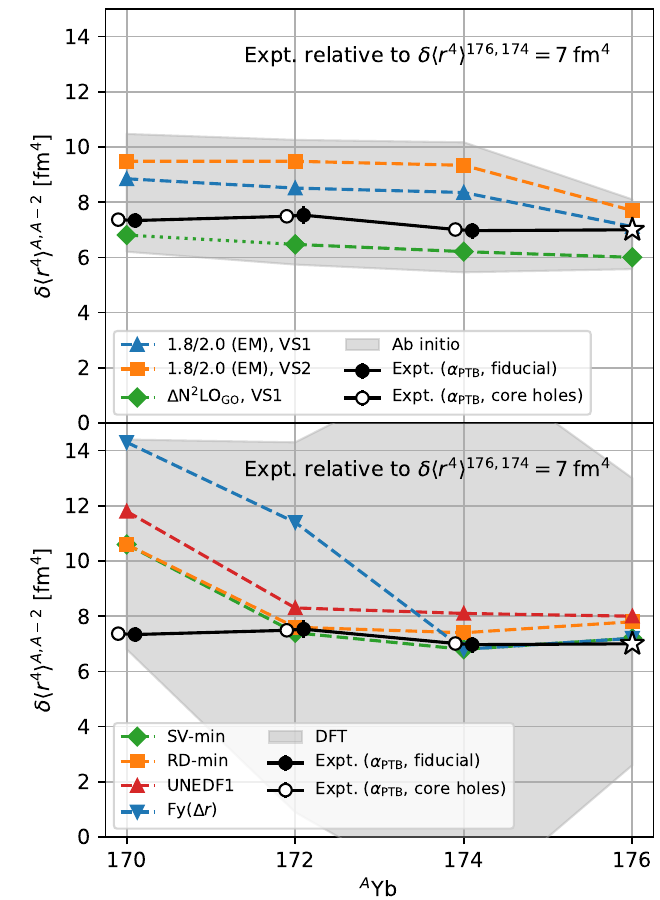} 
	\caption{Experimental $\drfour^{A,A-2}$ values relative to $\drfour^{176,174} = 7\:\fm^4$ extracted from isotope shifts from the $\alpha$ transitions using atomic theory (fiducial, core holes) are compared to nuclear theory predictions from our \textit{ab initio} calculations [top, 1.8/2.0~(EM), VS1 and VS2; $\Delta\text{N}^2\text{LO}_\text{GO}$, VS1] and from density functional theory calculations~\cite{Hur2022evidence} [bottom, SV-min, RD-min, UNEDF1, $\text{Fy}(\Delta r)$] for $A\in \{170,172,174,176\}$. The gray bands give estimated uncertainties of the theory results.
        }
	\label{fig:rp4comparison_dft}
\end{figure}

Next, we subtract the fit results from the $\hat{\nu}^{a, r}$ and obtain the residuals
\begin{align}
    \hat{\nu}_\tau^{a,r}- F_\tau D^{a,r} = G_\tau^{(4)} Q^{a,r}\,.
\end{align}
Making use of the calculated electronic coefficients $G_\tau^{(4)}$, which are listed in Tab.~\ref{tab:electronic-coefficients} we deduce three independent objects $Q^{a,r}$ per isotope-shift transition $\tau$. Fixing $r=(174,176)$ and $\tau=\alpha_{\mathrm{PTB}}$ and choosing a value $\drfour^r = -7\,\fm^4$, we are able to extract \drfour{} values for the remaining isotope pairs, given in Tab.~\ref{tab:exp_dr4}. We choose the reference value $\drfour^r = -7\,\fm^4$ informed by predictions from \textit{ab initio} nuclear structure theory and density functional theory shown in Fig.~\ref{fig:rp4comparison_dft}, which all predict this value to lie between 6--8~$\fm^4$. Beyond this chosen reference value, uncertainties and covariances are obtained by propagating uncertainties from experimental input ($\nu_\tau^a$, $w^a$, and $\drtwo^a$) and atomic theory ($G_\tau^{(4)}$) assuming independent normal distributions. The dominant source of uncertainty is $\drtwo^{(170,168)}$, and we also understand this uncertainty to lead to correlations in the fit in Eq.~\eqref{eq:dr4_fi_fit}, leading to the considerable anti-correlation between $\drfour^{(170,168)}$ and $\drfour^{(172,170)}$.

We note that the approach outlined here to extract information on \drfour{} from a King-plot nonlinearity is unable to give information on the absolute scale of \drfour{}. This is simply because we are considering only the nonlinearity, but \drfour{} also produces an isotope-shift contribution parallel to the linear field shift. We are unable to disentangle this contribution from the dominant field shift, so our approach is only able to make statements about how \drfour{} changes across the isotopic chain.

These values are compared with predictions for \textit{ab initio} nuclear structure theory and density functional theory in Fig.~\ref{fig:rp4comparison_dft}. Our extracted \drfour{} values show a weak decrease between $\drfour^{(172,170)}$ and $\drfour^{(174,172)}$ and otherwise flat trend in \drfour{} across the ytterbium isotopes studied here. Comparing this to the discussed nuclear structure calculations, we find that the flatter \drfour{} trends of \textit{ab initio} calculations more closely reproduce the extracted trends.

\begin{table}[th!]
    \centering
    \caption{
        Experimental $\drfour^{A,A-2}$ values relative to $\drfour^{176,174} = 7\:\fm^4$ extracted from isotope shifts from the $\alpha_\text{PTB}$ transitions using atomic theory with uncertainties and covariances propagated from input uncertainties.
    }	
    \begin{ruledtabular}
	\begin{tabular}{c|c|rrr}
		\multirow{3}{*}{\makecell{Isotope pair \\ $a$ = $(A,A')$}} 
            & \multirow{3}{*}{$\drfour{}^{a}$ [$\fm^4$]}
            & \multicolumn{3}{c}{cov($\drfour{}^{a}$, $\drfour{}^{b}$) [$\fm^8$]} \\
            & & \multicolumn{3}{c}{Isotope pair $b$ = $(A,A')$} \\
            & & (170,168) & (172,170) & (174,172) \\
		\hline
		(170,168)  & 7.33\,(27) &  \hphantom{$-$}0.076 & $-$0.088 & $-$0.010 \\
            (172,170)  & 7.53\,(32) & $-$0.088 &  0.104 &  \hphantom{$-$}0.006 \\
            (174,172)  & 6.97\,(28) & $-$0.010 &  0.006 &  \hphantom{$-$}0.079 \\
            (176,174)  & 7 (reference) &   \\
	\end{tabular}
    \end{ruledtabular}
	\label{tab:exp_dr4}
\end{table}

\section{Bounds on the new boson}
\label{app:boson_bounds}
\subsection{King-plot bounds}
The ytterbium spectroscopy bounds in Fig.~\ref{fig:exclusion_plot} in the main text were derived by constructing generalized King plots~\cite{Berengut:2020itu}, which require the isotope-shift and mass measurements of $n$ pairs of stable spinless isotopes on $n-1$ transitions to eliminate $n-2$ nuclear factors from the system of isotope-shift equations. In ytterbium, it is currently the number of suitable isotope pairs, $n=4$, that fixes the maximal dimension of the generalized King plot: Using the frequency measurements on 3 transitions, the charge radius variance $\drtwo$, as well as the leading King nonlinearity, induced by $\drfour$, can be eliminated.
The remaining King nonlinearity is used to evaluate the sensitivity of the isotope-shift data to the coupling $\alpha_{\mathrm{NP}}$ of the new boson to electrons and neutrons. This method is particularly useful since it only requires theoretical input in the form of the $D$-coefficients that appear in Eq.~\eqref{eq:A_sotopeshift_all} (see also Tab.~\ref{tab:mphi-Dcoeffs} and Fig.~\ref{fig:D_mphi}). 
Note that the subtraction of additional calculated SM effects such as $\drtwo^2$ from the remaining King nonlinearity would reduce the precision of the generalized King-plot bound, since this would introduce unknown theory uncertainties. Indeed, the main benefit of the generalized King plot method over direct comparisons between theory and experiment is that it is data-driven, taking advantage of the experimental precision.

\begin{figure}[th!]
   \centering
  \includegraphics[width=3.4in]{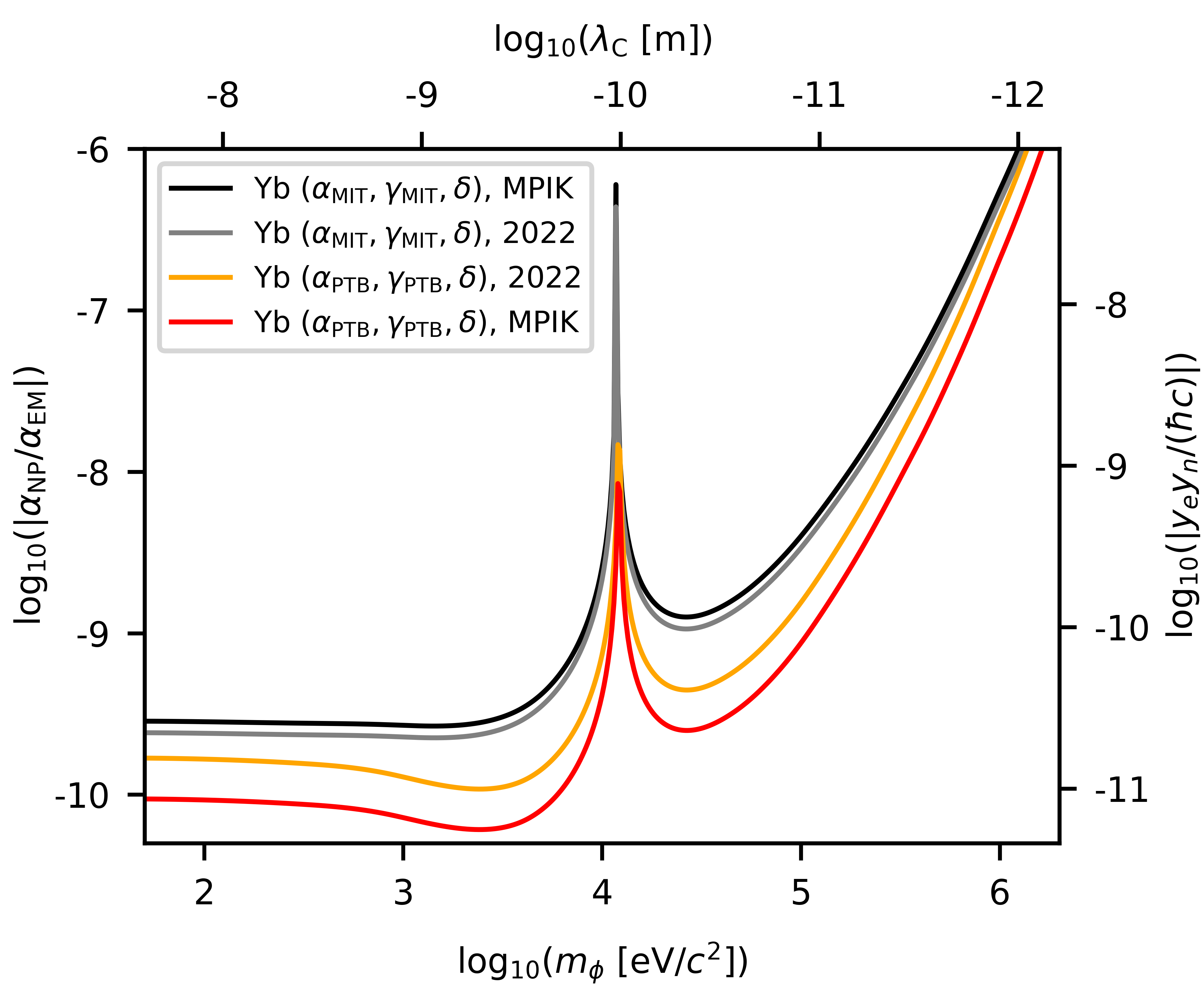} 
	\caption{Comparison of the generalized King-plot bounds for the transition frequencies $\alpha_\mathrm{MIT}, \gamma_\mathrm{MIT}$~\cite{Counts2020evidence,Hur2022evidence}, $\delta$~\cite{Ono2022observation} and $\alpha_\mathrm{PTB}$, $\gamma_\mathrm{PTB}$ presented in this work. The bounds employing the inverse mass differences stated in Ref.~\cite{Hur2022evidence} are labeled with ``2022'', the bounds employing the mass-ratio measurements presented in this work are labeled with ``MPIK''.  All bounds are computed using Eq.~\eqref{eq:def_alphaNP_bound}.
        }
	\label{fig:agd_curves_diff}
\end{figure}

The red curve in Fig.~\ref{fig:exclusion_plot} in the main text shows the bound on the new boson obtained by applying the generalized King-plot method to the frequency and mass measurements presented in this work, combined with the frequency measurements presented in Ref.~\cite{Ono2022observation}, and by computing
\begin{align}
    \max\left|\alpha_\mathrm{NP} \pm 2 \,\sigma_{\alpha_\mathrm{NP}}\right| / \alpha_\mathrm{EM} \,,
    \label{eq:def_alphaNP_bound}
\end{align}
where $\sigma_{\alpha_\mathrm{NP}}$ denotes the uncertainty on the new physics coupling $\alpha_\mathrm{NP}$ and is estimated using a simple Monte Carlo. (Simple uncertainty propagation leads to similar results.)
The red curve in Fig.~\ref{fig:exclusion_plot} in the main text thus corresponds to the absolute maximum of the upper and lower $2\sigma$ bounds on $\alpha_\mathrm{NP}/\alpha_\mathrm{EM}$. This choice is motivated by the fact that without full understanding of nuclear effects, King plots can only place bounds on new physics and thus should never exclude the case of zero new physics ($\alpha_\mathrm{NP}\to 0$). 
Although the sign of $\alpha_{\mathrm{NP}}=(-1)^{s+1}y_{\text{n}}y_{\text{e}}/(4\pi\hbar c)$ decides whether the new force is attractive or repulsive, showing the generalized King-plot bounds in the $(\alpha_\mathrm{NP}/\alpha_\mathrm{EM}, m_\phi)$ plane rather than the $(|\alpha_\mathrm{NP}/\alpha_\mathrm{EM}|, m_\phi)$ plane could lead to the false conclusion that the bounds on $\alpha_\mathrm{NP}$ become infinitely stringent in the case where both the upper and the lower bounds are on one side of $\alpha_\mathrm{NP}=0$. 
Let us simply note that the bound that we are showing in red in Fig.~\ref{fig:exclusion_plot} in the main text (and in Fig.~\ref{fig:agd_curves_diff}) would, in the $(\alpha_\mathrm{NP}/\alpha_\mathrm{EM}, m_\phi)$ plane, be negative for $m_\phi$ values below the peak at around 10~keV and positive for the plotted $m_\phi$ values above the peak.

In Fig.~\ref{fig:exclusion_plot} of the main text, our new bound is compared with  bounds constructed using the same methodology and mass-ratio measurements, but with Yb$^{+}$ isotope-shift data from Refs.~\cite{Hur2022evidence,Ono2022observation} (black, maroon curves), and with King-plot bounds obtained by applying the projection method of Ref.~\cite{Solaro2020improved}
to the Ca$^{+}$ data of Refs.~\cite{Solaro2020improved,Chang:2023teh} (cyan, blue curves). In the $(\alpha_\mathrm{NP}/\alpha_\mathrm{EM}, m_\phi)$ plane, the generalized King-plot bound derived for the set of transitions $\alpha_\mathrm{MIT}, \gamma_\mathrm{MIT}$, and $\delta$ (black curve) would have the opposite sign to the bound for $\alpha_\mathrm{PTB}, \gamma_\mathrm{PTB}$, and $\delta$ (red). This is due to the differences in the isotope-shift data $\nu_\alpha^{A,A+2}$ presented in Table~1 in the main text and provides yet another argument for interpreting the upper bound on $|\alpha_\mathrm{NP}|$ as a sensitivity of the respective dataset to the new boson rather than focusing on the sign of the bound. Indeed, the fact that for the datasets ($\alpha_\mathrm{PTB}, \gamma_\mathrm{PTB}$, $\delta$) and ($\alpha_\mathrm{MIT}, \gamma_\mathrm{MIT}$, $\delta$) the upper and the lower bounds on $\alpha_\mathrm{NP}$ are on the same side of $\alpha_\mathrm{NP}=0$ implies that a King-plot nonlinearity is dominating the system of isotope shifts, and should ideally be removed by constructing a higher-dimensional generalized King plot from a larger dataset.
The aforementioned offsets between the $\alpha_{\mathrm{MIT}}$ and $\alpha_{\mathrm{PTB}}$ measurements also affect the locations of the characteristic peaks in the generalized King-plot bounds. This explains the difference in the slopes of the black and red curves at the upper end of the plotted $m_\phi$ region. When these offsets are taken into account, these two slopes agree.

Figure~\ref{fig:agd_curves_diff} shows how the improved accuracy and precision of the frequency and mass-ratio measurements translate into sensitivity to the new physics coupling $\alpha_\mathrm{NP}$: Whereas higher precision leads to smaller uncertainties $\sigma_{\alpha_\mathrm{NP}}$ on the new physics coupling, and thus have the tendency to lead to stronger bounds on $|\alpha_\mathrm{NP}/\alpha_\mathrm{EM}|$, a shift in the central value of a mass or frequency measurement will generally lead to a shift of the central value $\alpha_\mathrm{NP}$ predicted by the generalized King-plot construction. This effect can be observed by comparing the black and the gray curves in Fig.~\ref{fig:agd_curves_diff}: Since the bound indicated in black was constructed using more precise mass measurements, one would naively expect it to be more stringent than the bound shown in gray that employs the same mass values as Ref.~\cite{Hur2022evidence}. 
However, the shift in the central value of $m_{168}$ between the old and the new mass-ratio measurements translates into an upward shift of the bound, which cannot be compensated by the increased experimental precision. 
In other words, the bound shown in gray appears to be more stringent than the bound shown in black since it underestimates a hitherto unknown systematic uncertainty.
Comparing the new bound presented by this work (red curve) against the gray and the orange curves, we see that both the improvement of the isotope-shift measurements and of the mass-ratio measurements have an impact on the final bound, and that the achieved sensitivity was only made possible by the combination of these improvements.

\begin{figure}[th!]
   \centering
  \includegraphics[width=3.4in]{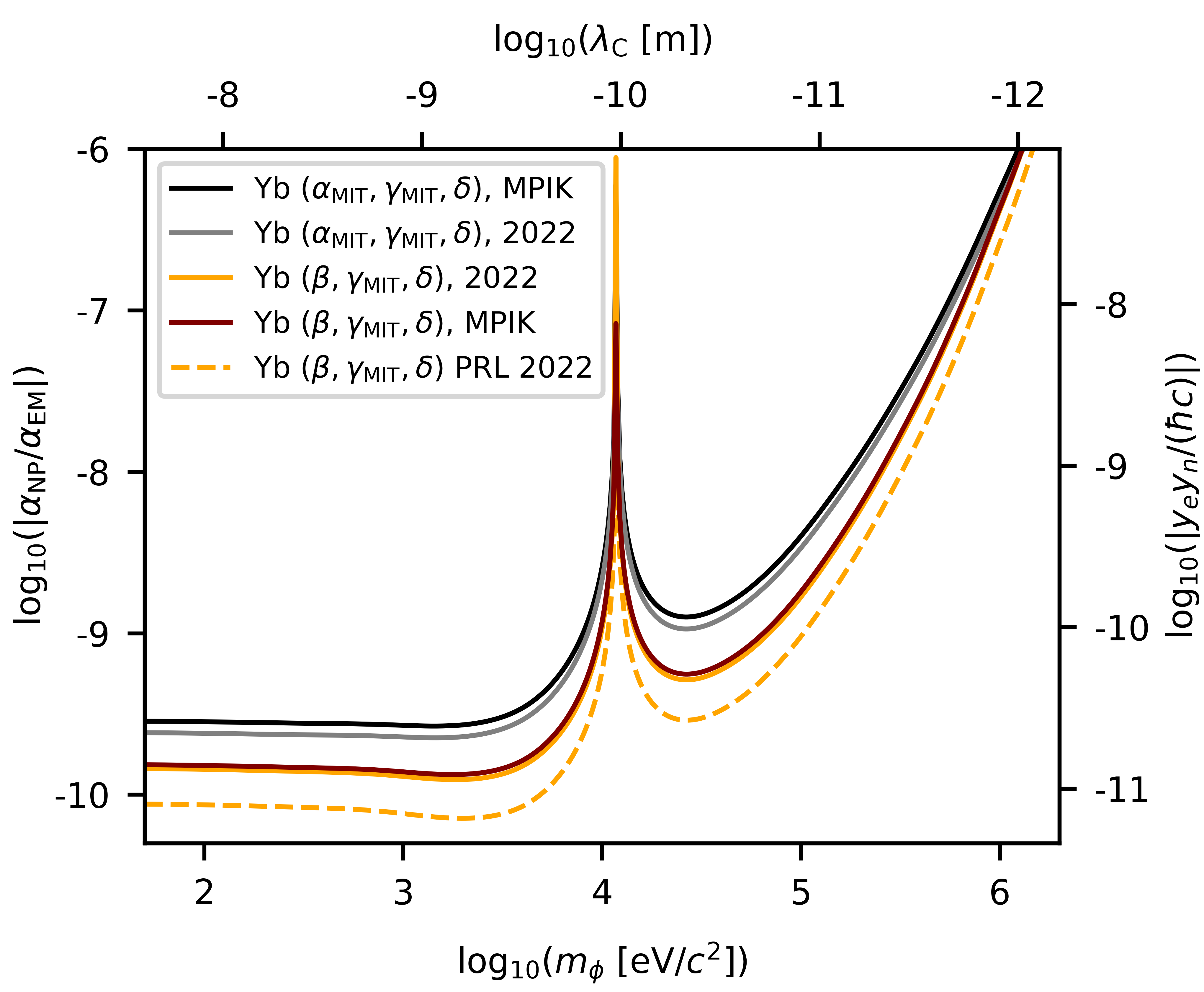} 
	\caption{Comparison of the King-plot bounds for the transition frequencies $\alpha_\mathrm{MIT}, \gamma_\mathrm{MIT}$~\cite{Counts2020evidence,Hur2022evidence}, $\beta$~\cite{Counts2020evidence} and $\delta$~\cite{Ono2022observation}. The bounds employing the inverse mass differences stated in Ref.~\cite{Hur2022evidence} are labeled with ``2022'', the bounds employing the mass-ratio measurements presented in this work are labeled with ``MPIK''. All curves apart from the orange dashed curve show generalized King-plot bounds. The orange dashed curve is taken from Ref.~\cite{Hur2022evidence}. All bounds are computed using Eq.~\eqref{eq:def_alphaNP_bound}.
        }
	\label{fig:MIT_curves_diff}
\end{figure}

Comparing the orange and maroon curves in Fig.~\ref{fig:MIT_curves_diff}, which show the generalized King-plot bounds obtained from the combination of the isotope-shift frequencies associated to the transitions $\beta$~\cite{Counts2020evidence}, $\gamma_\mathrm{MIT}$~\cite{Counts2020evidence,Hur2022evidence}, and $\delta$~\cite{Ono2022observation}, we observe a similar upward shift to that between the gray and the black curves. The dashed orange curve in Fig.~\ref{fig:MIT_curves_diff} corresponds to the absolute maximum of the upper and lower bounds on the new physics coupling presented in Ref.~\cite{Hur2022evidence}.
This is not a generalized King-plot bound, but a bound obtained from a combination of a fit and the theoretical input for the electronic coefficients $D_{\rho \sigma \tau}$, where $\rho$, $\sigma$ and $\tau$ are transition indices~\cite{Hur2022evidence}. Since the theoretical uncertainty on these triple-index electronic coefficients $D_{\rho \sigma \tau}$ is expected to be significantly larger than on the single-index electronic coefficients $D_\tau$, and since we aim to compare the experimental results on equal footing, we do not show this bound in Fig.~\ref{fig:exclusion_plot} in the main text but instead show the generalized King-plot bound computed using the same data.

\subsection{Competing bounds}
The competing $2\sigma$ bounds are shown as exclusion regions. The combination of the bound on $y_e$ from the measurement of the anomalous magnetic moment of the electron $(g-2)_e$~\cite{Hanneke:2008tm,Fan:2023gm2electronNorthwestern} and the bound on $y_n$~\cite{Frugiuele:2016rii} from neutron optics~\cite{Leeb:1992qf} and neutron scattering experiments~\cite{Nesvizhevsky:2007by,Pokotilovski:2006up,Barbieri:1975xy,Leeb:1992qf} is indicated in yellow.
For masses $m_\phi<10^2\,\mathrm{eV}/c^2$, fifth force searches via the Casimir effect provide particularly strong bounds (green)~\cite{Bordag:2001qi,Bordag:2009zz}.

In green we show the bounds from hydrogen-deuterium (HD) isotope shifts~\cite{Delaunay:2017dku} (see also Refs.~\cite{Potvliege:2023lvf,Jones:2019qny} for bounds on new physics from hydrogen and deuterium spectroscopy). For light elements, there are not sufficiently many isotopes to construct a King plot, but the precision of the theory predictions reaches the level of the experimental precision, such that a comparison of the two can be used to set a bound. This bound, however, depends sensitively on experimental input for the proton and deuteron charge radii. The hydrogen-deuterium bounds in Fig.~\ref{fig:exclusion_plot} in the main text are obtained using charge radii determined via electron scattering and muonic atom spectroscopy, respectively. 
The former leads to a weaker bound than the bound from $(g-2)_e$ times neutron scattering, and the latter could be affected by the new boson, if the latter couples to muons.

Beam dump experiments provide competitive bounds for $2m_e\lesssim m_\phi$. These are not shown in Fig.~\ref{fig:exclusion_plot} in the main text due to their dependence on the lifetime of the new boson $\phi$, which is a function of all couplings of $\phi$, not only of the couplings to electrons and neutrons. 

The blue-shaded areas in Fig.~\ref{fig:exclusion_plot} in the main text are disfavored by astrophysical considerations, such as bounds from star cooling in horizontal branch stars (globular cluster)~\cite{Grifols:1986fc,Grifols:1988fv} and from energy loss in the core of the supernova SN1987a~\cite{Raffelt:2012sp}, which is subject to an $\mathcal{O}(1)$ uncertainty~\cite{Blum:2016afe}. Complementary bounds from laboratory experiments would provide valuable contributions to the search for new bosons also in these regions of parameter space. 

Finally, the green dashed lines in Fig.~\ref{fig:exclusion_plot} in the main text show the bounds on $y_e$ from the anomalous magnetic moment of the electron $(g-2)_e$~\cite{Hanneke:2008tm,Fan:2023gm2electronNorthwestern} multiplied by the bounds on $y_n$ from the kaon decay $K\to \pi+\mathrm{invisible}$, which can be estimated using the bounds on the new boson couplings to quarks and the first coefficient of the QCD $\beta$-function~\cite{Knapen:2017xzo}.
These bounds are highly dependent on the flavor structure of the couplings of the new boson $\phi$ to the quarks. For illustration, Fig.~\ref{fig:exclusion_plot} in the main text shows the bound on $\alpha_{\mathrm{NP}}/\alpha_{\mathrm{EM}}$ in the presence of a coupling to top-quarks, $y_t$, and the bound in presence of a coupling to up-quarks, $y_u$. They differ by approximately seven orders of magnitude.

This leaves open regions of parameter space where King plots have the potential of providing the most competitive bounds on $\phi$ that are agnostic both of its decay widths into dark matter and of the flavor structure of its couplings to quarks.
In the future, the King-plot bounds can be further improved by increasing the dimension of the generalized King plots. This can be achieved by expanding the experimental program to isotopes with long half-lives (e.g., ${}^{166}$Yb) or by moving to new atomic systems such as Sn~\cite{leibrandt2022prospects} or Xe~\cite{rehbehn23prl}, both of which have seven, spinless stable isotopes and suitable clock transitions.


\end{document}